# Modeling and Theoretical Design on Next-Generation Lithium Metal Batteries

Yanchen Fan[1†], Xiang Chen[2†], Dominik Legut[3], Qianfan Zhang[*1]

[1]School of Materials Science and Engineering, Beihang University, Beijing 100191, P. R. China.

[2]Department of Chemical Engineering, Beijing Key Laboratory of Green Chemical Reaction Engineering and Technology, Tsinghua University, Beijing 100084, P. R. China

[3]IT4Innovations Center, VSB-Technical University of Ostrava, 17. listopadu 15, CZ-708 33 Ostrava, Czech Republic

[†] These authors contributed equally to this work.
*Corresponding authors: qianfan@buaa.edu.cn.

## Acknowledgement:

Q.F.Z. was supported by National Natural Science Foundation of China (11404017), Technology Foundation for Selected Overseas Chinese Scholar, Ministry of Human Resources and Social Security of China and National Key Research and Development Program of China (No. 2017YFB0702100)

# Modeling and Theoretical Design on Next-Generation Lithium Metal Batteries

## Abstract

Rechargeable lithium metal batteries (LMBs) with an ultrahigh theoretical energy density have attracted more and more attentions for their crucial applications to portable electronic devices, electric vehicles, and smart grids. However, the implementation of LMBs in practice are facing numerous challenges, such as low Coulombic efficiency, poor cycling performance, and complicated interfacial reactions. First-principles calculations has been a powerful technique in battery researches, in terms of modeling the structure and properties of specific electrode materials, understanding the charge/discharge mechanism at the atomic scale, and delivering a rational design strategy for electrode materials as well as electrolytes. In this review, theoretical studies on lithium–sulfur (Li–S) batteries, lithium–oxygen (Li–O$_2$) batteries, lithium metal anode, and solid-state electrolytes (SSEs) are summarized. The first two chapters mainly focus on issues concerning with cathode in Li–S and Li–O$_2$ batteries. The theoretical researches on the Li metal anode and SSEs are particularly reviewed. The current challenges and further developments of LMBs are prospected from a theoretical viewpoint.

# 1. Introduction

Secondary lithium ion batteries (LIBs) are crucial for a wide range of applications in human daily life, including electric vehicles, grid storage, and advanced portable devices [1, 2]. However, the current LIB techniques cannot satisfy the energy demands in the future due to their theoretical energy density limits. Rechargeable lithium metal batteries (LMBs) based on multiple-electron reactions rather than ion-intercalation are expected to handle with this challenging issue for their higher theoretical energy densities [3-5]. Specially, lithium–sulfur (Li–S) batteries and lithium–oxygen (Li–$O_2$) batteries are strongly considered as most promising candidates for next-generation energy storage devices for their ultrahigh theoretical energy densities (non-aqueous Li–$O_2$ battery: 3505 Wh $kg^{-1}$; Li–S battery: 2600 Wh $kg^{-1}$) [6-11]. Furthermore, solid-state LMBs can even deliver a much higher energy density in practice due to the decreasing of electrolyte mass. The introduction of SSEs is also expected to fundamentally solve the fatal defect on the flammability and limited electrochemical stability of organic liquid electrolytes as well as to stabilize the Li metal anodes [5, 12].

Unfortunately, the application of the rechargeable LMBs still faces numerous challenging issues, such as low Coulombic efficiency, poor cycling performance, and complicated interfacial reactions [4, 13]. Each specific LMB also possesses respective issues. For example, the application of Li–S batteries is seriously hindered by some intrinsic issues: (1) the migration of soluble polysulfides driven by both concentration gradients and electric field, which is famous as shuttle effects, causes the side reaction between Li metal and polysulfides; (2) the significant volume fluctuation around 80% due to lower density of $Li_2S$ (1.66 g $cm^{-3}$) than that of sulfur (α-phase, 2.07 g $cm^{-3}$) induces a distinct change to electrode structure; and (3) the low electrical conductivity of sulfur and solid product cause the low utilization of active materials [8, 14-18]. Besides, the research about Li–$O_2$ batteries is still in primary stage, the actual energy density is far less than its theoretical energy density, and the capacity fading is also a critical trouble [19]. Particularly, Li metal anodes are troubled by the uncontrollable growth of lithium dendrites, which can induce a low Coulombic efficiency and severe safety hazards, and SSEs are challenged by the low ion conductivity and high interfacial resistance [20].

Tremendous efforts have been devote to handle these intractable issues to promote the practical demonstration of LMBs with a very high energy density. Among various techniques, theoretical modeling based on quantum mechanics, such as density functional theory (DFT) calculations and Hartree–Fock methods [21-23], plays an important role in these researches. Specially, they can provide a fundamental understanding of the physical and chemical properties of battery materials, such as geometrical structure, electronic interaction, ion diffusivity, phase transition, and kinetic lithiation process, a fruitful insight into the charge/discharge mechanism, and thus a rational design strategy of LMBs [24-26]. For example, Wang *et al.* [27] calculated the interfacial lithium storage capacity of $Li_2O$ through the DFT calculations and explained the additional capacity of the metal oxide observed during the experiments. Iddir *et al.* [28] investigated the phase stability of $xLi_2MnO_3$ (1−x) $LiMO_2$ according to the formation energy and the migration energy of Co vacancies based on the first-principles calculations.

Despite many impactful and insightful reviews, in which Li–S batteries [8, 9, 14-17, 29], Li–$O_2$ batteries [30-32], Li metal anodes [4, 33, 34], and SSEs [35-40] have been well summarized, none of them particularly focuses on the theoretical progresses toward next-generation LMBs. The scope of this review is to provide an overview of major theoretical progresses on LMBs recently, in terms of Li–S batteries, Li–$O_2$ batteries, Li metal anodes, and SSEs. The first two chapters mainly focus on issues concerning with cathode about Li–S and Li–$O_2$ batteries. Theoretical researches about the Li metal anodes and SSEs are particularly discussed in the next two chapters due to their specializations, respectively. For each section, an introduction of current research condition and the major concerned theoretical issues are provided at first, followed by representative works concerning with these issues as well as a summarization of the remaining problems. In the end of the review, current challenges and further developments of LMBs are summarized and prospected from a theoretical viewpoint.

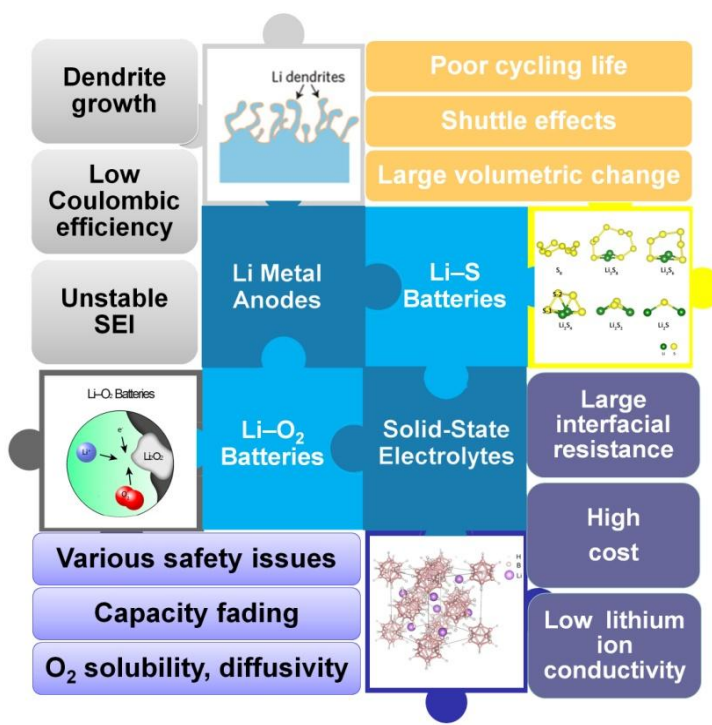

**Figure 1. Modeling of next-generation LMBs and the corresponding crucial challenges [41-44].**

## 2. Modeling of Li–S batteries

Li–S batteries are typical and promising energy storage devices for a multitude of emerging applications. The sulfur cathode with a specific capacity of 1672 mAh $g^{-1}$ can deliver a high energy density of 2600 Wh $kg^{-1}$ when match a Li metal anode (Figure 2a), which is 5 times larger than that of conventional LIBs based on Li metal oxide cathode and graphite anode [45-49]. Furthermore, the low cost and low toxicity provide sulfur cathode many advantages for commercial applications. Despite the considerable advantages, the implementation of Li–S batteries has been hindered by the tremendous obstacles, including the poor cycle life, low Coulombic efficiency, and low utilization of active material [50, 51]. Specifically, the soluble long-chain lithium polysulfides (LiPSs) can be reduced to $Li_2S$ when diffuse to anode surface and react with Li metal, inducing the loss of active species and the passivation of the anode (Figure 2b). Besides, both sulfur and discharge product, such as $Li_2S$, have low ionic/electronic conductivities, endowing a low utilization of sulfur. Even if the sulfur is reduced to $Li_2S$ completely, the cathode will experiences a high volumetric change as large as 80% due to the huge change of gravimetric density from S to $Li_2S$. Therefore, a composite cathode is crucial to design advanced Li–S batteries.

The success of Li–S batteries lies on regulating the dissolution of LiPSs and enhancing the conductivity of the sulfur cathode. A suitable sulfur host with specific component and topological structure is required to maintain the active sulfur species to achieve a stable long-term cycle life. Besides, the charge/discharge mechanism based on multi-electron reactions in Li–S batteries, which is totally different from the Li-ion intercalation chemistry in LIBs, arises more requirements for cathode materials, such as promoting the conversion among LiPSs. Theoretical calculation is a powerful and effective tool on screening promising sulfur hosts and revealing the interaction between LiPSs and various hosts at the atomic scale. In this section, we summarized the major developments of sulfur cathode, concentrating on the interaction between LiPSs and sulfur hosts as well as the key parameters to achieve a rational design of sulfur hosts.

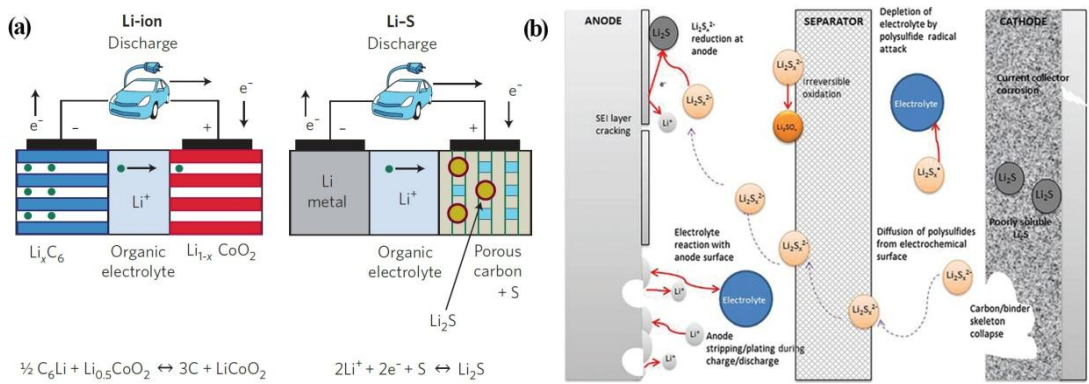

Figure 2. (a) The basic reactions schematic of Li-ion batteries and Li–S batteries [19]; (b) The dissolution of LiPSs degradation mechanisms [53].

## 2.1. Thermodynamic properties

An ideal sulfur cathode is supposed to provide sufficient room for sulfur volumetric expansion, strong affinity toward LiPSs, and conductive surface for $Li_2S$ deposition. Previously, carbon-based materials with high electrical conductivity are commonly used. However, there is still a considerable capacity decay during long-term cycling due to the weak interaction between carbon materials and polysulfides and as-caused dissolution of polysulfides into electrolytes during the charge and discharge processes. Therefore, the key of a rational design of emerging sulfur cathodes is to introduce a strong interaction between polysulfides and sulfur hosts [54-58].

Polymer binders, which can improve the interaction between polysulfides and carbon substrate as well as current collectors, are supposed to resist the shuttle effects [59]. Choosing an appropriate polymer binder is very critical to maintain the active species in cathode and achieve an excellent cycling performance. In 2013, Zheng *et al.* [60] reported the modification method of the sulfur cathode by adding amphiphilic poly-vinylpyrrolidone (PVP), which has strong binding with carbon surface due to the strong thermodynamic driving force in eliminating the hydrophobic interface. The polar functional group in PVP can provide a large binding energy toward LiPSs simultaneously, thus delivering a stable long-term cycling with a decay smaller than 3% over the first 100 cycles. This phenomenon was further investigated with first-principles calculations to understand the intrinsic chemistry. Common carbon frameworks, modeled with graphene herein, can only provide a small binding energy smaller than 0.3 eV toward $Li_2S$ and LiS molecules. However, PVP, modeled with N-methyl-2-pyrrolidone (NMP) as their share the same functional group, can induce a much larger binding energy of 0.66–1.29 eV toward $Li_2S$ and LiS molecules, respectively. The strong chemical binding between LiPSs and NMP originated from the polar functional group in polymer, which is different from the physical confinement induced by routine carbon materials. The introduction of strong chemical binding is considered as the key factor to prevent the shuttle of soluble LiPSs, providing a new insight for the rational design of the sulfur cathodes based on first-principles calculations.

To compare the effects of different polymer binders with different functional groups in anchoring LiPSs, a variety of polymers based on the framework of vinyl polymers $-(CH_2-CHR)_n-$ has been considered (Figure 3a) [61]. Among various functional groups, the >C=O contained functional groups, including esters, ketones, and amides, induce the largest binding energy toward $Li_2S$. Other functional groups exhibit a relatively smaller binding energy, especially for the halogenated ones (F-, Br-, and I- contained functional groups). The geometry analyses of O-containing PVP and F-containing PVDF absorbed with $Li_2S$ and LiS are further provided in Figure 3b and 3c, confirming the strong chemical binding induced by oxygen atom in PVP. This work shows the great significance of screening high-performance polymer binders for sulfur cathodes based on first-principles calculations [14].

Many other polymers which can provide high affinity toward LiPSs are also explored. For example, conductive polymers, such as polyaniline (PANI), polypyrrole (PPY), and poly (3,4-ethylenedioxythiophene) (PEDOT), are applied to sulfur cathode as they can induce strong confinement effect to LiPSs and simultaneously enhance the electrical conductivity of the composite cathode [62]. For example, Ma *et al.* reported a high-performance sulfur–carbon cathode based on tethering polyethylenimine (PEI) polymers with large numbers of amine groups on functionalized multiwall carbon nanotubes [63]. Chen *et al.* synthesized a binder with a specific hyper-branched network structure and abundant polar amino groups, which can provide a strong affinity of LiPSs and thus greatly improve the cycle performance [64]. Up to now, developing an efficient binder is still a good choice to resist the shuttle effects and achieve a stable long-term cycling performance for Li–S batteries.

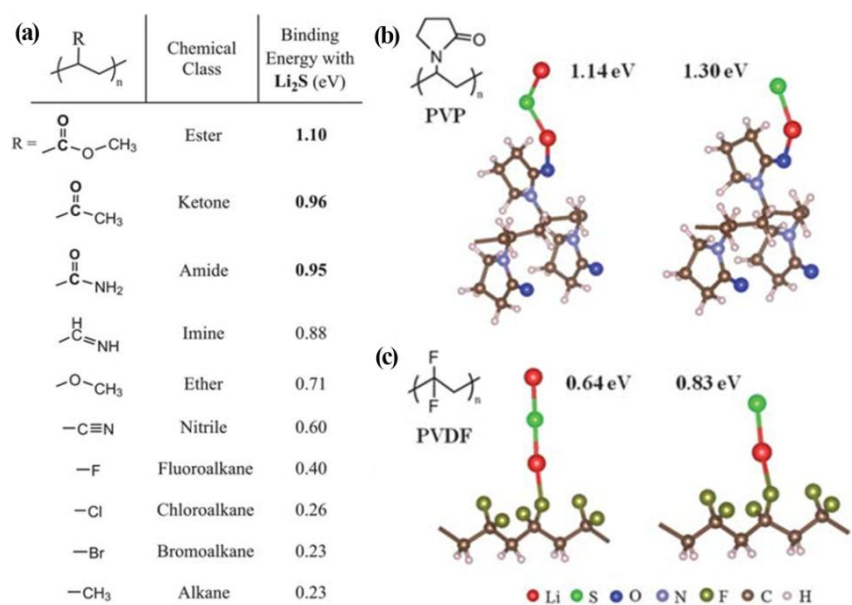

**Figure 3.** (a) The binding energy o various functional groups toward Li$_2$S. (b) The most stable configuration and corresponding binding energies of PVP absorbed with Li$_2$S and LiS. (c) The most stable configuration and corresponding binding energies of PVDF absorbed with Li$_2$S and LiS [61].

Another alternative to anchor LiPSs is to introduce inorganic additives into sulfur cathodes. The requirements for choosing additives follow similar rules with that of polymer binders, in terms of binding strength, conductivity, and stability. It is commonly believed that the anchoring effect by such inorganic components can be stronger than organic polymer, which is supposed to achieve a better performance for Li–S batteries. Various oxides, such as Mg$_{0.6}$Ni$_{0.4}$O [65], Al$_2$O$_3$ [66], tin-doped indium oxide [67], TiO$_2$ [68], and Ti$_4$O$_7$ [69, 70], are firstly considered to immobilize polysulfides. More recently, metal sulfides, such as TiS$_2$, VS$_2$, ZrS$_2$ [71, 72], NiS [73], MoS$_3$ [74], CoS$_2$ [75], and Co$_9$S$_8$ [76], as well as hydroxides [77] and nitrides, are applied to sulfur cathode due to their higher electrical conductivity.

First-principles calculations are also applied to simulate the anchoring strength of these metallic components toward LiPSs, mainly focusing on the binding energy and charge transfer. For polymer binder, a stronger binding energy can generally induce a better performance. However, the similar situation seems different for the metallic additives. On one hand, metal sulfides are supposed to deliver a weaker binding effect toward LiPSs than metal oxides. However, sulfur cathode containing metal-sulfide additives exhibits a battery performance as well as, even better, that containing metal-oxide additives. On the other hand, a metallic components with ultrahigh binding energy toward LiPSs can cause the irreversible decomposition of LiPSs, reducing the cycling performance of the batteries. All of these indicate that the mechanism, relating the anchoring strength with practical battery performance, is very complicated, and more detailed and comprehensive investigations are required to gain more insights to achieve a rational design of a composite sulfur cathode with metallic additives.

In order to compare the anchoring effects of different metallic additives, Zhang *et al.* conducted a systematic investigation on layered materials as sulfur hosts, including various metal oxides, sulfides, and chlorides [42]. The binding energy are summarized in Figure 4a. For the adsorption of $S_8$, these materials exhibit a similar binding energy of 0.75–0.85 eV, which is dominated by physical van der Waals (vdW) interaction. For LiPSs, different materials induce distinct binding energy, to which chemical interaction contributes the major part. Generally, oxides, sulfides, and chlorides can be regarded as strong-, moderate-, and weak-anchoring substrates, with the binding energies in ranges of 2.0–4.2, 0.8–2.0, and 0.4–0.8 eV, respectively. According to the absorption configuration and charge transfer analyses (Figure 4b), the anchoring effect originates from the distinct charge transfer between LiPSs and anchoring material that electron density between them is increased, which is defined as chemical interaction. More specifically, the strong chemical interaction is realized by a Li–O bond, which also weakens the Li–S bond in LiPSs. Furthermore, a too strong anchoring can even lead to the decomposition of LiPSs, such as the condition of $V_2O_5$ or $MoO_3$ (Figure 4c). The decomposed LiPSs is very hard to be re-oxidized to sulfur and thus has a negative effect on the battery performance. In contrast, sulfur hosts with moderate anchoring effects can strike a balance between resisting shuttle effect and maintaining a relative chemical stability of LiPSs, which are supposed to be a good choice for advanced sulfur cathodes.

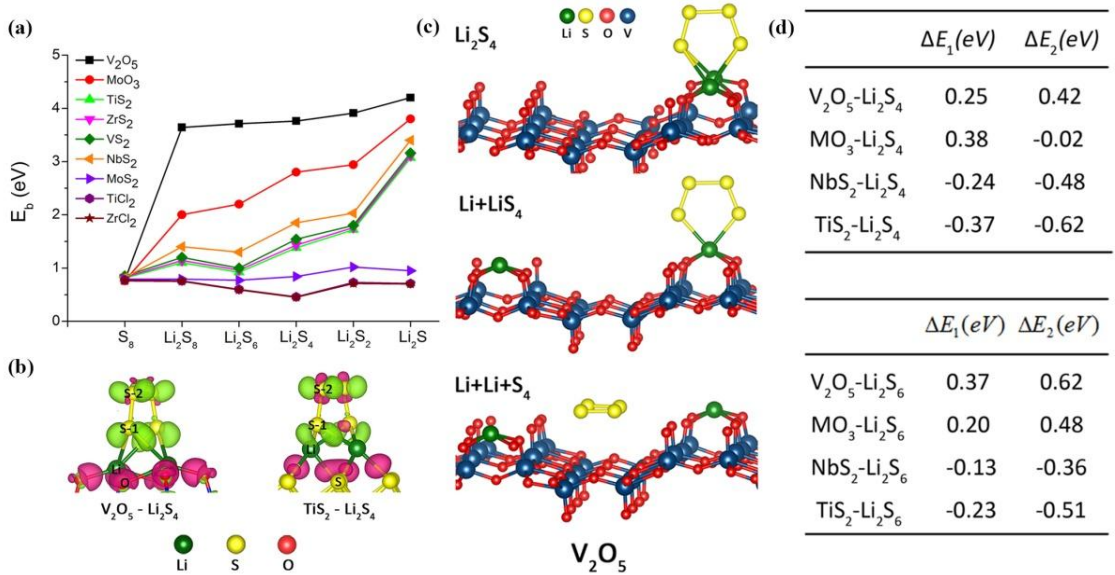

**Figure 4 (a) The binding energies between metallic components and sulfur species (such as $S_8$, $Li_2S_8$ $Li_2S_6$, $Li_2S_4$, $Li_2S_2$, $Li_2S$). (b) Charge transfers between $V_2O_5$/$TiS_2$ and $Li_2S_4$. (c) Geometrical configurations for $Li_2S_4$ (first), Li + $LiS_4$ (second), Li + Li + $S_4$ (third) clusters adsorbed on $V_2O_5$ surface. (d) Energy differences between intact $Li_2S_4$ and destructed $Li_2S_4$'s adsorption on $V_2O_5$. Moreover, $\Delta E_1 = E(Li_2S_4 + V_2O_5) − E(Li + LiS_4 + V_2O_5)$, and $\Delta E_2 = E(Li_2S_4 + V_2O_5) − E(Li + Li + S_4 + V_2O_5)$, respectively [42].**

The importance of physical interaction in sulfur cathode systems was noticed and first-principles calculations based on vdW correction were widely used since then. More and more anchors have been explored based on the theoretical predictions, and they are proved to efficiently optimize the performance of sulfur cathode systems. Among them, the sulfide and oxide anchors are good candidates, such as $Co_9S_8$ [76], $CoS_2$ [75], $MnO_2$ [78], α-$Fe_2O_3$ [79] and ZnO [80]. Many study also focus on the layered structured materials, especially for simulations due to the perfect surface without chemical binding breaking and the appearance of dangling bond of layered structured materials, which results in a reasonable binding strength independent on the surface chosen for modeling. For example, Lei *et al.* used polar $WS_2$ nanosheets as freestanding electrodes and performed the theoretical calculations, revealing the $WS_2$ nanosheets with a moderate anchoring to polysulfides [81]. Zhao *et al.* systematically exploited the anchoring effects of monolayered MXenes, $Ti_2CO_2$ and $Ti_3C_2O_2$ [82]. The interaction between Li in LiPSs and O atoms in MXenes is remarkable but not too strong, so the intactness of the LPSs can be well maintained.

Although the interaction between LiPSs and anchors have been investigated, a deep insight into the intrinsic chemistry of this interaction is still lacking. Inspired by the concept of hydrogen bond, Hou *et al.* focused on the Li bond chemistry in Li–S batteries through sophisticated quantum chemical calculations [83]. The electron-rich donor (e.g., pyridine nitrogen) in N-doped graphene interacts with the Li atom in LiPSs, forming a Li–N interaction, namely lithium bond. The chemical nature of lithium bond was further confirmed as a dipole–dipole interaction by structure, binding energy, dipole, and charge transfer analyses (Figure 5). The chemical shift of lithium polysulfide in the $^7$Li NMR spectrum was used to quantitatively determine the strength of the lithium bond both experimentally and theoretically. Further electrochemical test demonstrates this theory. This work highlights the importance of Li bond chemistry in Li–S battery and provides a rational design strategy to carbon based sulfur cathode. While for the first-row metal sulfides, sulfur-binding plays a more important role in anchoring LiPSs than lithium-binding, which is different from the carbon-based sulfur hosts [84].

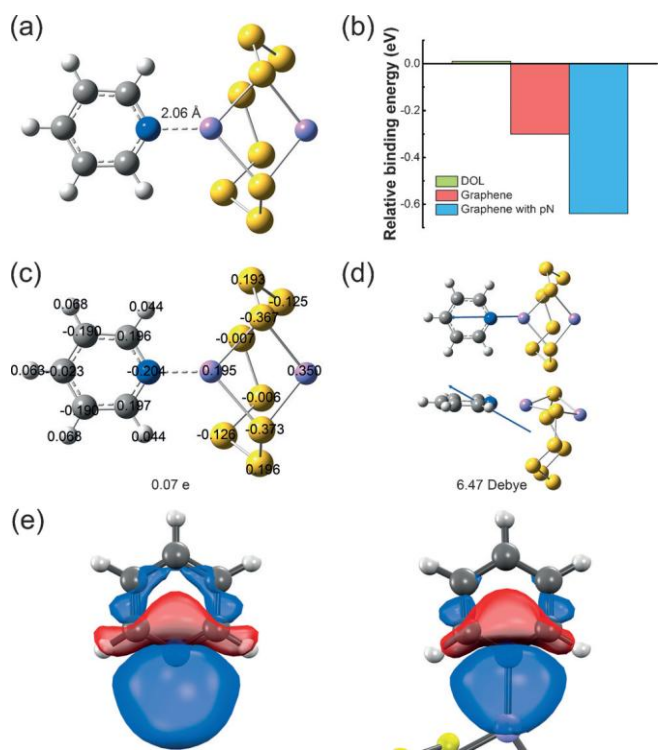

**Figure 5.** (a) Optimized geometry of $Li_2S_8$ binding to pyridine. (b) Binding energies between $Li_2S_8$ and DOL, graphene, or graphene doped with pyridinic nitrogen. (c) Charge transfer between $Li_2S_8$ and pyridine. (d) Dipole analysis of $Li_2S_8$–pyridine cluster. (e) Natural bond analysis of $Li_2S_8$ before and after interacting with pyridine [83].

## 2.2. Kinetic properties

Besides thermodynamic properties, kinetic properties also play an important role in determining the performance of Li–S batteries. The kinetic properties are specially critical in understanding the chemical and electrochemical conversions among various sulfur species, including sulfur molecules, ions, and even radicals. Generally, the exploration of the kinetic properties is of much more difficulties than that in investigating the thermodynamic properties. Directly investigating the charge/discharge process is still very challenging and researchers mainly focus on the key steps in a working Li–S cell, such as the diffusion of the $Li^+$ ion and the oxidation of the $Li_2S$.

Five kinds of metal oxide, including $CeO_2$, $Al_2O_3$, $La_2O_3$, MgO, and CaO, were selected as the modeling system to understand the relationship between adsorption and diffusion of LiPSs on nonconductive sulfur hosts [85]. Although all of the oxides can induce a very strong anchoring effect toward LiPSs, the lithium diffusivities can be quite different, which makes a distinct difference on the actual battery performance. A lower surface diffusion barrier leads to a faster Li diffusion and higher deposition efficiency of sulfide species on electrodes. Hence, the selection of nonconductive additives for sulfur hosts should balance the surface adsorption and diffusion of LiPSs.

To prove the catalytic oxidation function of metallic additives in sulfur cathode, a series of metal sulfides are investigated through a systematic experimental study in combination with first-principles calculations [86]. Compared with carbon materials, $Ni_3S_2$, $SnS_2$, FeS, $VS_2$, $TiS_2$ and $CoS_2$-contained cathodes have a higher specific capacity, lower over-potential and better cycle stability (Figure 6b), which is explained by the catalytic function of metal sulfides. The decomposition barrier of $Li_2S$ molecule on various substrates was selected as the descriptor of determining the catalytic function of metal sulfides in promoting the oxidation of $Li_2S$. Specifically, the decomposition barrier on $VS_2$, $TiS_2$, and $SnS_2$ surfaces are around 0.3 eV, which is much smaller than that on graphene surface, around 1.8 eV, which agrees with the experimental initial charge voltage profile analyses (Figure 6c). The strong interaction between the metal sulfides and $Li_2S/Li_2S_x$ can reduce the energy barrier and change the reaction pathway (Figure 6d–j), which can promote the transport of $Li^+$ ions, control the surface precipitation of $Li_2S$, accelerate the surface-mediated redox process, and improve the Li–S battery performance. Hence, the catalytic function is a very important parameter when choosing sulfur hosts for advanced Li–S batteries.

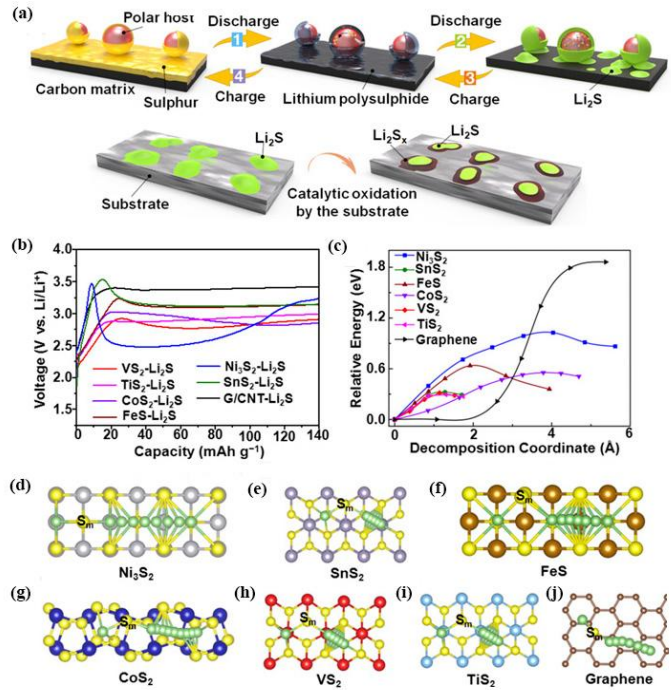

**Figure 6. (a)** Schematic illustration of the sulfur conversion process and the $Li_2S$ catalytic oxidation on the surface of the routine carbon substrate (top) and various metal sulfides (bottom). **(b)** First cycle charge voltage profiles of $Ni_3S_2$−$Li_2S$, $SnS_2$–$Li_2S$, FeS−$Li_2S$, $CoS_2$–$Li_2S$, $VS_2$−$Li_2S$, $TiS_2$−$Li_2S$, and G/CNT−$Li_2S$ electrodes. **(c)** Energy profiles for the decomposition of $Li_2S$ cluster on $Ni_3S_2$, $SnS_2$, FeS, $CoS_2$, $VS_2$, $TiS_2$, and graphene. The schematic representations of the corresponding decomposition pathways for **(d)** $Ni_3S_2$, **(e)** $SnS_2$, **(f)** FeS, **(g)** $CoS_2$, **(h)** $VS_2$, **(i)** $TiS_2$, and **(j)** graphene [10].

Up to now, people have achieved a deep insight into Li–S battery cathode, which greatly improve the theoretical understanding of the interaction between LiPSs and sulfur hosts at the atomic scale. However, there are still some problems remained to be solved. Firstly, small clusters or single molecules of $Li_2S_n$ are previously used to model the interaction between LiPSs and sulfur anchors, which is too simplistic and deviates from the real condition in a working Li–S cell. Secondly, the charge/discharge mechanism on cathode are rarely investigated due to the complexity of $Li_2S_n$ conversions, which is not only associated with the Li-ion intercalation but also involved with the overall structure changes of sulfur species. Thirdly, the multiphase transition from solid sulfur to soluble LiPSs ($S_8$, $Li_2S_8$, $Li_2S_6$ and $Li_2S_4$) and finally to solid state of $Li_2S_2$ and $Li_2S$ still remains tremendous difficulties to be clear. Therefore, much more efforts, including theoretical and experimental works as well as their combinations, should be devoted to these remained challenging issues.

# 3. Modeling of Li–O$_2$ batteries

Rechargeable Li–O$_2$ batteries have been widely studied as a large-scale energy storage technique since 1996 due to their ultrahigh theoretical energy density [87]. In Li–O$_2$ batteries, the Li$^+$ ions react with the reduced oxygen in cathode side, while the occurred reaction are different for non-aqueous and aqueous electrolytes (Figure 7). Currently, the practical demonstration of Li–O$_2$ batteries is still facing enormous challenges, including various safety issues and relative low practical energy density. Therefore, tremendous efforts have been devoted to Li–O$_2$ batteries, in terms improving the low energy conversion efficiency, exploring the complex charge/discharge mechanism, stabilizing the Li metal anode, resisting the corrosion of the lithium, and developing high-efficient cathode catalysts [88].

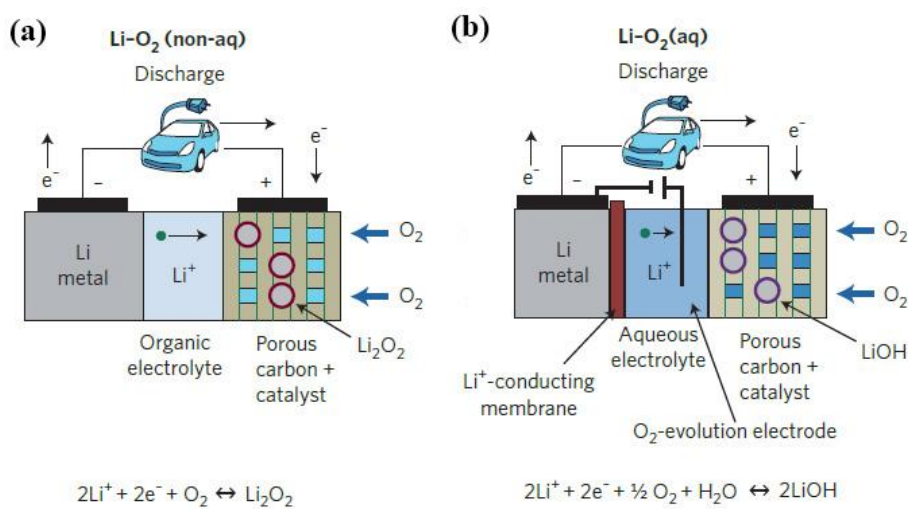

**Figure 7. The basic reactions schematic of two common Li–O$_2$ batteries [52].**

Li–O$_2$ battery cathode plays an important role in achieving an advanced battery performance in terms of providing the oxygen diffusion channels, catalyzing the reduction of oxygen, accommodating the discharge products, and regulating the morphology of the reaction products. Ideal cathode materials should be conducive to the accumulation of discharge products and the facilitation of the reaction between lithium and oxygen [89]. The rational material design is important on achieving a high-efficient cathode, which requires a fundamental understanding of the electrodes and reaction mechanism. In the following text, we will summarize the recent progresses on the theoretical study of Li–O$_2$ battery cathode, from both aspect of Li-O product and the design of catalyst.

## 3.1. Discharge product

During discharge processing, Li$^+$ ions gradually react with reduced oxygen and produce solid state Li$_2$O$_2$ on the cathode. Both thermodynamic and kinetic properties relating to this product is critical for optimizing the performance, the electronic structure, and the ion/electron transport. Besides, the nucleation growth are the most important issue that need to be clearly clarified in advance.

The Li$^+$ ion and electron diffusions are closely related to the thermodynamic and kinetic properties of Li$_2$O$_2$. It has been proved that Li$_2$O$_2$ is intrinsic insulator with a very large band gap [90, 91]. The electron transport should be realized through the hopping of polarons, which is generated by the confinement of electrons around the O–O bond. Such polaron model is the main origin of the weak transport effect in the intrinsic Li$_2$O$_2$ [92, 93]. However, the situation can be quite different on the surface. Radin *et al.* [94-96] systematically studied the electronic structure of surfaces for both Li$_2$O$_2$ and Li$_2$O and forty oriented surfaces were considered. The oxygen-rich <0001> and <1−100> surfaces can be very stable for Li$_2$O$_2$, while the <111> surface with both Li and O is stable for Li$_2$O. Further study showed that Li$_2$O$_2$ is metallic and magnetic, while Li$_2$O is non-metallic and non-magnetic, which can reveal different electrochemical performance during initial charging state, and confirm the conductivity of Li$_2$O$_2$ on the surface.

The charge/discharge overpotential is also closely related to the discharge product, Li$_2$O$_2$. Hummelshøj *et al.* [97] focused further on facet-dependent overpotential of Li$_2$O$_2$ on various facets, terminations, and sites (terrace, steps, and kinks) of a Li$_2$O$_2$ surface. Very low overpotentials (<0.2 V) for both discharge and charge are found at different facet sites, agreeing well with the experiment observation. Beyond surface system, it is also commonly believed that the ion or electron transport can be carried out through the delocalized hole by Li vacancy in bulk region when Li$_2$O$_2$ deviated from the stoichiometric ratio [98]. During the initial discharge process, the nonstoichiometric Li$_{2-x}$O$_2$ can be produced, which represent a thermodynamically favorite process and results in a fast kinetic step. Therefore, the overpotential for discharge is not so large [92, 93]. For the charge process, in contrast, the decomposition of Li and O$_2$ from Li$_2$O$_2$ should be much more difficult, and the overpotential increases accordingly.

The kinetic process of Li$_2$O$_2$ growth is also very important, and multiscale modeling was demonstrated as a useful tool to study the reaction and diffusion mechanism in Li–O$_2$ battery [99-101]. Recently, Yin *et al.* [102] have a detailed study on the nucleation mechanism of Li$_2$O$_2$ by applying the universal multiscale model combining nucleation theory, reaction kinetics, and mass transport. According to the evolution of the calculated cell potential, the discharge processes follows four steps: (I) LiO$_2$ accumulation, (II) Li$_2$O$_2$ nucleation, (III) Li$_2$O$_2$ growth, and (IV) sudden death (Figure 8b). During stage I, O$_2$ is reduced and LiO$_2$ is formed and accumulated, leading to a decrease of the potential (Figure 8c). During stage II, nucleation initiates with the decrease of $LiO_2$ concentration as soon as that the excess energy of the super-saturation can overcome the nucleation barrier and $Li_2O_2$ generates, leading to the increase of the potential and the decrease of the nucleation rate (Figure 8d). During stage III, the $Li_2O_2$ particles gradually grow. On one hand, the growth of active surface area for $LiO_2$ adsorption induce the decrease of $LiO_2$ concentration and the increase of the potential. On the other hand, the decrease on active surface area for $O_2$ reduction leads to an increase of the decrease of the cell potential. Such combined effect results in a bending shape of the potential profile. During stage IV, the impact from the electrode passivation dominates, and the remaining free surface of the electrode deactivate quickly along with the sharp drop of the cell potential, causing "sudden death". According to the particle dimension distribution (Figure 8e), the main part of the capacity comes from the particles with a radius around 90 nm, which are mainly formed during stage II, while the peak around small particles with the radius of 12 nm are mainly formed during stage IV. This simulation agreed with the experimental results well, and deeply revealed the nucleation and growth mechanism of $Li_2O_2$ on the cathode.

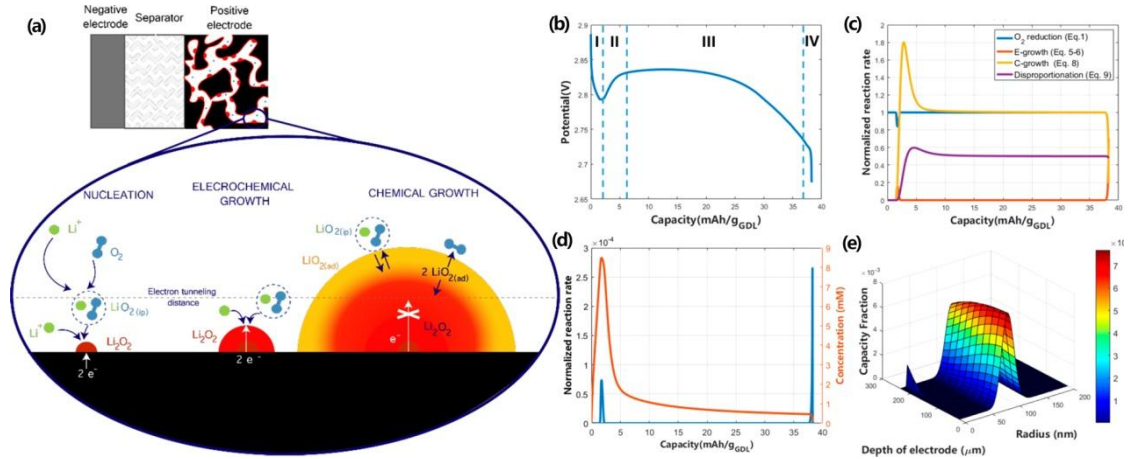

**Figure 8. (a) Schematic illustration of the discharge process in Li−$O_2$ batteries. (b) The calculated potential profile during the discharge process. (c) The reaction rate of $O_2$ reduction, growth of $Li_2O_2$ together with reduction of $LiO_2$ and $LiO_2$, growth of $Li_2O_2$ together with the adsorption of $LiO_2$, and disproportionation of $LiO_2$. (d) The evolution of the nucleation rate and $LiO_2$ concentration. (e) The particle dimension distribution at the end of the discharge process [102].**

Due to pure electronic conductivity and high overpotential, replacing $Li_2O_2$ by other discharge product (such as $LiO_2$, which is the intermediate superoxide in the reaction and has been synthesized in experiment) is considered as an alternative choice to overcome these problems. Through first-principles simulations [103-105], especially accurate HSE method [104], $LiO_2$ was confirmed as an insulator with a relative large band gap (~3.7 eV) yet much lower charging overpotentials than that of $Li_2O_2$. Based on these work, Li *et al.* [106] systematically studied the charge transport mechanism in LiO$_2$. According to the band structure, both excess electrons and holes are confined around the oxygen dimer, while small polarons that can transfer by hopping among neighboring lattice sites will form. Further calculation on concentrations and mobility of the charge carriers exhibited that the charge transportation in LiO$_2$ is governed by the migration of hole polarons and positively charged oxygen dimer vacancies along *c*-axis (with barriers as low as 0.25 and 0.32 eV, respectively). As a result, the electronic conductivity associated with such polaron hopping ($3 \times 10^{-12}$ S cm$^{-1}$) is eight orders of magnitude higher than that in Li$_2$O$_2$, and can solve the key problem in Li–O$_2$ batteries.

## 3.2. Catalysts

The practical applications of Li–O$_2$ batteries are primarily prevented by the sluggish kinetics of discharge/charge reactions, causing serious problems such as high overpotential, low current density, unstable electrodes, and electrolyte decomposition. Developing high-performance catalysts in both oxygen reduction reaction (ORR, 2Li$^+$ + O$_2$ + 2e$^-$→Li$_2$O$_2$) and oxygen evolution reaction (OER, Li$_2$O$_2$→2Li$^+$+O$_2$ + 2e$^-$) is critical to realize a high-performance Li–O$_2$ battery. The catalyst for ORR can increase the discharging voltage and reduce the discharging overpotential, while the catalyst for OER can reduce both the charging voltage and charging overpotential. The first-principles calculations can be used to reveal catalytic reaction mechanism and develop novel high efficient catalysts, which become increasingly important. Many simulations and theoretical researches focus on modeling the adsorption of Li-O products and intermediate on catalyst, based on which the realistic reaction process can be identified and the key factors that determine the reaction properties can be clarified.

The noble metal was initially found to be with good catalytic activity. Chen *et al.* [107] theoretically studied the deposition of Li$_2$O$_2$ on both Pt and Au metal by applying non-equilibrium Green function based DFT computation, and found that the generation of Li$^+$ vacancy can lower the anti-bonding orbit and help for the loss of electron and the oxidation process. Therefore, the noble metal can induce a very strong electronic interaction with Li$_2$O$_2$, and exhibit the catalytic functions for both ORR and OER. Dathar *et al.* [108] comparably studied several kinds of metals, and found that the activity of Au, Ag, Pt, Pd, Ir, and Ru for ORR forms a volcano-like profile with respect to the adsorption energy of oxygen (Figure 9a). The theoretical analysis indicated that too strong oxygen−metal interaction (Ru and Ir) diminishes the capacity of oxygen binding with Li, whereas too weak oxygen−metal interaction (Au and Ag) hinders O$_2$ activation and oxygen−lithium bonding. Therefore, the intermediate oxygen−metal interaction strengths on Pt and Pd are best. The "volcano-like" behavior between ORR potential and oxygen adsorption energy was also found by Lu *et al.* [43] (Figure 9b). Combining experiment and simulation, the ORR activity was found to be in the order of Pd > Pt > Ru≈Au>carbon.

After that, people found that the alloy by noble metal and earlier transition metal can be more efficient on catalyzing the $Li_2O_2$. Kim *et al.* [109] found $Pt_3Co$ nanoparticles can serve as an effective OER catalyst in Li–$O_2$ batteries. The adsorption energies of Li and $LiO_2$ are suggested as critical descriptors of catalytic activity. $Pt_3Co$ showed a high binding strength for $Li^+$ ion and lower binding strength for intermediate $LiO_2$ comparing with both pure Pt and pure Co, which leads a high catalytic activity of $Pt_3Co$. They further studied PtCo in the following work [110], and found that PtCo exhibits over-potentials of 0.19 and 0.20 V for ORR and OER, respectively, remarkably lower than the overpotentials of both pure Pt (1.02 and 1.62 V) and $Pt_3Co$ (1.02 and 1.13 V). Such improvement can also be attributed to the strong Li yet weak $LiO_2$ adsorptions on the surface of PtCo. The ORR and OER overpotentials can even significantly decrease 0.34 and 0.83 V when replacing Co with Ti to produce $Pt_3Ti$ catalysts [111].

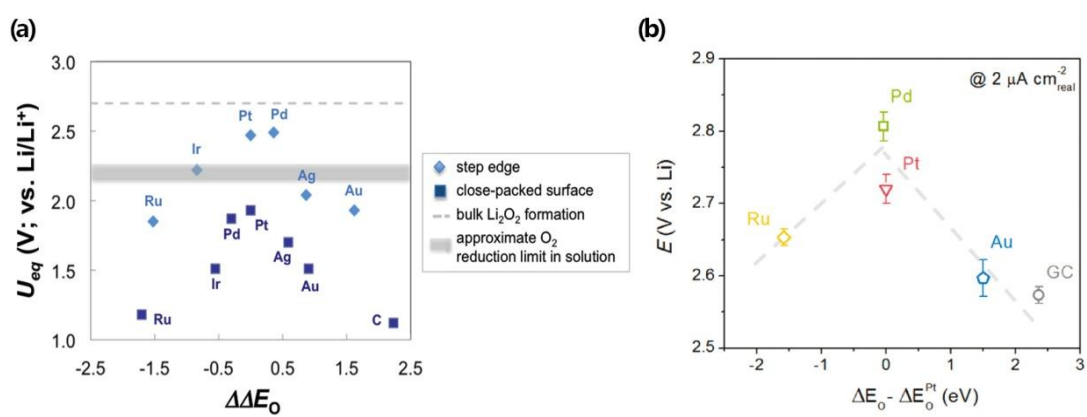

**Figure 9.** (a) Equilibrium potential ($U_{eq}$) of the first $e^-$ transfer step in the ORR reaction on six kinds of metals plotted against the adsorption energy of O atom relative to that on Pt metal, on both close-packed and step edge surfaces, respectively [108]. (b) ORR potentials at a current of 2 μA $cm^{-2}$ as a function of calculated oxygen adsorption energy $\Delta E_O$ (per oxygen atom relative to an atom in the gas phase), relative to that on Pt metal [43].

Comparing with expensive noble metals, transition metal components with a lower price as well as an excellent catalytic activity attract more and more interests recently. For example, $MnO_2$ is viewed as a promising OER and ORR catalyst for Li–$O_2$ batteries [112][113-118]. Specifically, the large (2×2) tunnel spaces in α-$MnO_2$, which is formed by the $MnO_6$ octahedrons, is very important for containing $Li_xO_y$ and providing the space for the reversible reaction and "$Li_xO_y$ storage" [112]. The formation of "$Li_xO_y$-$MnO_2$" structure can act as the intermediaries during discharge and charge reactions. Besides, the $Mn^{3+}$ ion with lower valence state is conductive to the reduction reaction. Ling [118] and Zhou *et al.* [115] further suggested the catalytic mechanism and the key steps from both aspects of $Li_2O_2$ decomposition and $O_2$ dissociation on the surface of α-$MnO_2$. Liu *et al.* [116] studied the catalytic effect of δ-$MnO_2$ nano-sheet. The $Li_2O_2$ film can be uniformly deposited onto δ-$MnO_2$ and the $Li_2O_2$/$MnO_2$ interface can serve as the good electrical conductor. However, the further oxidation of $Li_2O_2$ will be hindered by the high overpotential of 1.21 V according to their meso-scale simulations. Mellan *et al.* [117] studied the adsorption of Li and O atoms on β-$MnO_2$. The $O_2$ prefers to firstly deposit on the surface of $MnO_2$ through the dissociation and formed an O-rich surface, which promotes the deposition of $Li^+$ ion and leads to a low energy barrier for the initial reduction of oxygen.

Despite $MnO_2$, various other oxides have been investigated. For example, Li *et al.* [119] study the catalytic effect of $CeO_2$. The adsorption behavior of oxygen and Li on $CeO_2$ surfaces and reaction pathways for both ORR and OER were identified. Zhao *et al.* [120] presented novel $Co_2CrO_4$ nano-spheres as the electro-catalysts, revealing the key factors and steps involved in the $Li_2O_2$ formation and decomposition. Shi *et al.* [121] took an effective and widely used catalyst, rutile $RuO_2$, as an example to studied its catalytic mechanism in Li–$O_2$ batteries. The calculation results confirmed that the proposed catalytic scenario is both thermodynamically and kinetically viable. Zhu *et al.* [122] studied the oxygen evolution reaction mechanism of $Li_2O_2$ supported on active surfaces of transition-metal compounds, including the oxides such as TiO, NiO, MnO, ZnO and $Co_3O_4$, together with other materials like $Mo_2C$, TiN and $TiCl_3$. The key descriptor that determines the catalytic activity were fully discussed that the surface acidity of catalyst plays a critical role. Besides, $O_2$ evolution and $Li^+$ desorption energies show a linear and volcano relationships with surface acidity, respectively. Materials with an appropriate surface acidity can reduce the charging voltage and activation barrier. Accordingly, $Co_3O_4$, $Mo_2C$, TiC, and TiN are predicted to have a relatively high catalytic activity, which was also demonstrated by experiments.

Graphene-based catalysts were also found to be very efficient on promoting the reaction in Li–$O_2$ battery cathode, especially by introducing dopant or defect. Kang *et al.* [123] demonstrated that the N-doping graphene possesses catalytic activities toward both ORR and OER processes. Besides, Lee [124] and Tu *et al.* [125] also found the similar phenomenon and studied the adsorption enhancement mechanism of Li oxides onto N-doped graphene. Yun *et al.* [126] carried out a comparative study of the catalytic role of various defective site. The computational analysis demonstrated that the graphitic-N site is much more effective serving as catalytic media when comparing with either pyridinic- or pyrrolic-N site. Wu *et al.* [127] demonstrated that B-doping graphene exhibits much stronger interactions with Li-O clusters, which can activates Li−O bonds to decompose $Li_2O_2$ more effectively. Then, Ren *et al.* [128] made thermodynamic calculations to systematically study the catalytic activity of a series of X-doped graphene (X = B, N, Al, Si, and P) materials. The calculated OER pathways and their corresponding charge voltages for the doped graphene are shown as Figure 10a, while those for B, P-co-doped graphene was calculated and the energy profile of the reaction path is displayed in Figure 10b. Among these materials, P-doped graphene exhibited the highest catalytic activity on reducing the charge voltage by 0.25 V, while B-doped graphene showed the highest catalytic activity in decreasing the oxygen evolution barrier by 0.12 eV. Therefore, the catalytic effect of P-doped graphene is mainly activating the $Li^+$-desorption, while B-doped graphene exhibited a high catalytic activity in reducing $O_2$ evolution barrier.

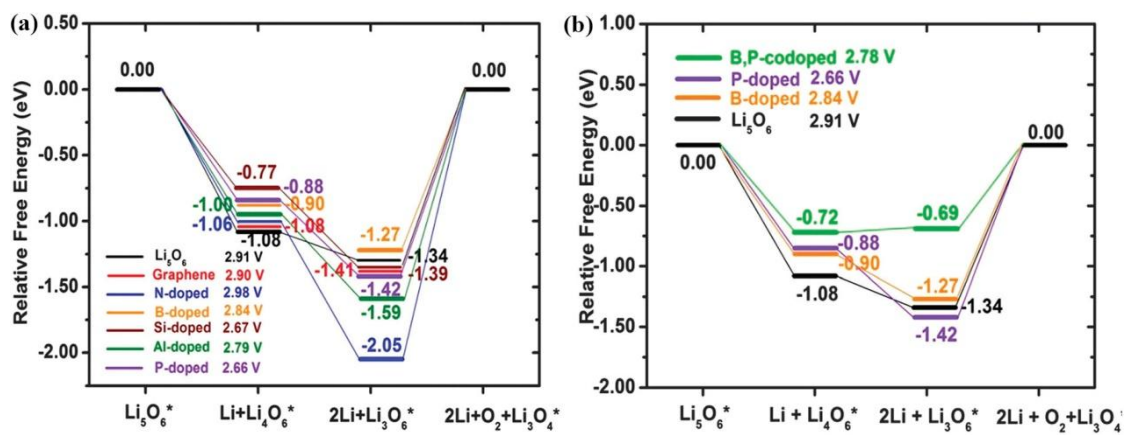

**Figure 10. (a) The thermodynamic OER pathways of various doped graphene. (b) The thermodynamic OER pathways of B, P-co-doped graphene [128].**

During recent years, many other kinds of catalysts are also explored. Beyond graphene-based materials, several other types of layer-structured materials with high catalytic activity are also theoretically predicted, such as h-BN supported by Ni (111) [129], silicene [130], and siligraphenes [131]. It is also found that the encapsulation of active Fe nanorods into N-doped carbon nanotubes (Fe@NCNTs) can be useful for the long-life Li–$O_2$ batteries [132]. The charge transfer between the interface of Fe nanorods and CNT walls and the active N dopants can enhance the chemisorption and subsequent dissociation of $O_2$ molecules. Jang *et al.* [133] found that the RuSe$_x$ nanoparticles exhibited a dramatic improvement in both the reaction kinetics and cyclic stability comparing with $RuO_2$ oxide, and the simulation showed that the $Li_2O_2$ product is more stable on RuSe$_x$ comparing with $RuO_2$, and induce the lower discharge voltage. Zhang *et al.* [134] discovered that the soluble bio-catalyst of coenzyme $Q_{10}$ (CoQ$_{10}$) can efficaciously drive solution phase formation of $Li_2O_2$. The calculated results indicated that the $Li_2O_2$ product can induce stronger interaction with CoQ$_{10}$ than with carbon cathode. As results, the small-dimensioned $Li_2O_2$ species can form in the electrolyte solution, and more time is required to nucleate into large $Li_2O_2$ particles on cathode surface, avoiding the "sudden death" phenomenon by the passivation of insulating $Li_2O_2$.

Currently, the development of Li–$O_2$ batteries still faces tremendous challenges. According to the simulation works mentioned above, theoretical methods can be served as a powerful tool to explore the fundamental reaction mechanism and optimize a high-performance catalyst for oxygen cathode. Similar with Li–S batteries, many questions remain far from clear, including the electronic interaction picture and the detailed reaction mechanism. Simultaneously, the reaction between lithium and other components in air, such as $N_2$, $CO_2$ and $H_2O$, remains one of the most serious problems in Li–$O_2$ batteries. How to inhibit these reactions can be a valuable topic for theoretical researchers.

# 4. Modeling of Li metal anode

Li metal possesses the lowest negative electrochemical potential (−3.04 V vs. standard hydrogen electrode), low density (0.534 g cm$^{-3}$), and an ultrahigh theoretical specific capacity of 3860 mAh g$^{-1}$, which is about ten times larger than that of conventional graphite (372 mAh g$^{-1}$) anode [135, 136]. With the emergence of post Li ion batteries, Li metal anode is thus considered as the "Holy Grail" in energy storage and the ultimate solution for next-generation high-energy-density batteries. However, the practical applications of Li metal anode have been facing numerous challenges over the past 40 years, such as the large volume expansion of anode during charge/discharge cycling, complicated interfacial reactions between Li metal anode and electrolyte, the uncontrollable growth of lithium dendrites, and "dead lithium" (Figure 11) [137-139].

The growth of dendrite and unstable solid electrolyte interphase (SEI) are two decisive factors facing Li metal anodes. Tremendous strategies have been proposed to suppress the growth of lithium dendrites and stabilize Li metal anode–electrolyte interphase, including the modification of electrolytes [140-144], regulation of solid electrolyte interphase [145-149], and construction of three-dimensional lithiophilic scaffolds [150-155]. However, the fatal issues of Li metal anode are still far away from solved due to the lack of a basic understanding of the fundamental chemistry in Li metal anodes, such as the mechanism of the interfacial interactions between anodes and electrolytes and Li metal deposition. Theoretical calculation is a powerful tool to fundamentally clarify these intractable issues. Three questions are mainly concentrated on from a theoretical viewpoint: (1) the interfacial reactions between anode/SEI and liquid electrolyte; (2) the properties of solid electrolyte interphase and protective layers; (3) the nucleation of Li$^+$ ion on the anode surface.

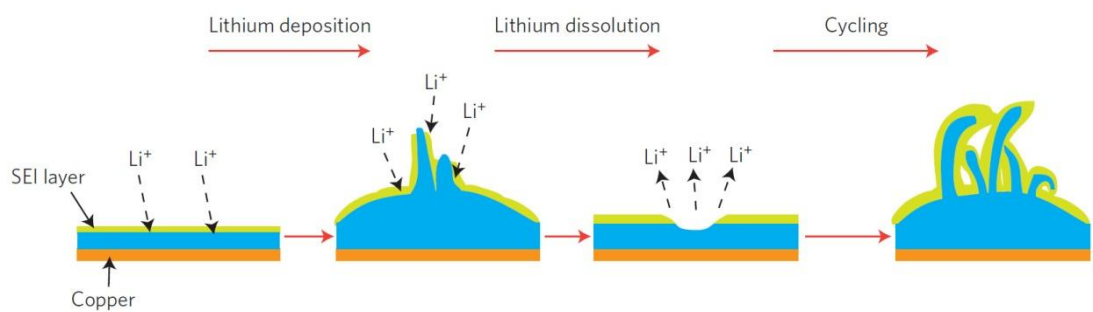

**Figure 11. A thin film of SEI layer forms quickly on the surface of deposited Li (blue). Volumetric changes during the Li deposition process can easily break the SEI layer. This behavior leads to ramified growth of Li dendrites and rapid consumption of the electrolytes [156].**

## 4.1. Interfacial reactions

Routine organic electrolytes spontaneously decompose on the anode due to the high reactivity of Li metal, which leads to capacity degradation and serious safety issues. Simultaneously, gas evolution is accompanied and produce various flammable gasses, including alkanes, alkene, and carbon monoxides. Understanding the mechanism of the interfacial reactions between electrolytes and Li metal anodes is very critical to construct a stable SEI and thus a stable Li metal anode.

Understanding the mechanisms of the interfacial reactions between electrolytes and anodes is prior to build a stable interface. Chen *et al.* [157] investigated the gas evolution mechanism based on DOL and DME electrolytes. The formation of lithium-ion–solvent complexes was confirmed as the key factor promoting the decomposition of electrolytes on lithium anodes. Comparing to DOL, DME exhibits a relative better stability towards Li metal anode, which was confirmed by both theoretical calculations and experiments. Camacho-Forero *et al.* [158] also investigated the effect of electron-rich environments on the decomposition mechanism of electrolyte species in 1,2-dimethoxyethane (DME) solvent. They characterized the thermodynamics and kinetics of the most relevant electrolyte decomposition reactions. DME decomposition reactions predicted from the MD methods were found to be thermodynamically favorable under exposure to Li atoms and/or by reactions with salt fragments. Recently, high-concentration DME electrolyte has been demonstrated with special functions to stabilize lithium metal anode [159]. Most anions exist in coordination condition and all $Li^+$ are solvated in high-concentration electrolytes according to molecular dynamics simulations.

Electrolyte additives play an important role in the interfacial reactions and formation of SEI. Camacho-Forero *et al.* [160-163] first studied the interfacial reactions between Li metal and electrolytes with additives such as VC (vinylene carbonate). Recently, fluoroethylene carbonate (FEC), which is one of the most promising additives for robust SEI formation, was used to form a LiF-rich SEI and stabilize the Li metal anode. Zhang *et al.* [140] investigated the kinetics of the formation of the LiF-rich coating(Figure 12a–d), and it is found that FEC additive is prior to be decomposed on Li metal anode than EC/DEC solvents due to the relatively lower the lowest unoccupied molecular orbital (LUMO) level. Therefore, the compact and stable FEC-induced coating layer is beneficial to obtain a uniform morphology of Li deposits (Figure 12e), guaranteeing the efficient operation of LMB with long cycle life.

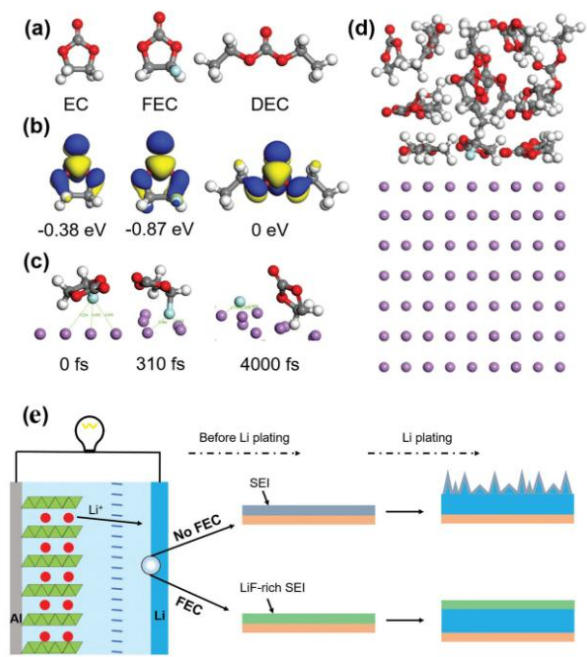

**Figure 12.** (a) Molecular structures of EC, FEC, and DEC. (b) Visual LUMO and corresponding relative energy of EC, FEC, and DEC. (c) *ab initio* molecular dynamics model. (d) Complete sequence of FEC molecule decomposed on Li anode. (e) Schematic illustration of the effect of FEC additives on Li metal anode [140].

Various types of lithium salts are found to be very efficient to form a high-quality protective film. Camacho-Forero *et al.* [164] studied the interfacial reactions between Li metal and DME/LiTFSI or DOL/LiTFSI electrolytes, respectively. LiTFSI reacted with Li metal anode immediately and the decomposition reaction can rapidly generate LiF protective film. However, DME and DOL are relatively stable without any decomposition reaction observed. When considering the concentration of LiTFSI salt in electrolytes, LiTFSI mainly exists in trans-structure in low concentration electrolyte yet cis-structure in high concentration electrolyte [165]. This cis-trans structure of Li salts change also affects the structure of the complex solvation networks. Thus, the structure can affect the metal interface reaction behavior on Li metal surface. Chen *et al*. [166] found that the Li$^+$ ion-solvent complexes greatly promoted the decomposition and gas evolution of the electrolytes through first-principles calculations in combination with in situ optical microscopic observations. This mechanism is applicable to various ion–solvent systems, including PC, DOL, DME, and TEGDME electrolytes on lithium or sodium metal anodes.

For Li metal anode in Li–S batteries, polysulfides can also serve as the coating layer that prevent the excessive consumption of Li metal [167-169]. Camacho-Forero *et al.* [164] studied the interfacial reaction between the polysulfide in the electrolyte and Li metal anode. It was found that $Li_2S_8$, including ring and linear geometries, was completely reduced to $Li_2S$ on the surface of Li metal. Soon after that, Liu *et al.* [170] theoretically modeled the $Li_2S$/Li interface and elucidated the nucleation and growth of a $Li_2S$ film on the anode surface due to long-chain polysulfide decomposition during battery operation. They found that $Li_2S$ tends to generate disordered interlayer on Li (110) surface, whereas tends to produce a perfect $Li_2S$ (111) surface on Li (111) surface.

## 4.2. SEI

Most organic solvents can react with reactive Li metal and the SEI film can be formed during the initial charging/discharging processes, which exhibits electrically insulating and ionically conductive nature [171]. Constructing a stable and efficient SEI is one of the most effective approaches to inhibit the growth of lithium dendrites, in which understanding the formation mechanism, composition, and stability of SEI is very important [167, 172, 173]. Theoretical simulations are supposed to provide a fruitful insight into these challenging issues, which is of tremendous difficulties from an experimental exploration. Specifically, the surface chemistry, surface morphology, electrochemical property, and dynamic characteristics of SEI can be well understood through first-principles calculations.

The $Li^+$ ion transport mechanism in SEI is one the most fundamental issues. Shi *et al.* developed a multiscale theoretical methodology to reveal it [174]. They first set up mesoscale diffusion equations formulated from a two-layer/two-mechanism model. Then the equations including pore diffusion in the outer and knock-off diffusion in the inner were confirmed by related experiments [175]. Then, a lot of essential theoretical issues, including $Li^+$ ion conductivity and electron transport, have been investigated based on above method. Some important conclusions have been drawn. (1) The dominant $Li^+$ ion diffusion carriers in $Li_2CO_3$ over a voltage range (0–4.4 V) were identified and it depended on the voltage. Despite the similar SEI components in cathode and anode sides, the dominant $Li^+$ ion diffusion carriers are different that excess $Li^+$ ion interstitials dominate below 0.98 V while the $Li^+$ ion vacancies become dominant above 3.98 V [176]. (2) They first pointed out that two different transport mechanisms of $Li^+$ ion existing in the two-layer structured SEI. The porous outer layer can conduct both $Li^+$ and anions, while the dense inner layer facilitate only can transport $Li^+$ ion [177]. (3) By calculating electron tunneling barriers from Li metal through various insulating SEI components, they also found that the formation and continuous growth of SEI layer are both responsible for the irreversible capacity loss of batteries in the batteries cycles [178]. These works provided a deep insight into the formation, structure, and properties of various SEIs.

The characteristics for various SEIs have been clearly understood in terms of electronic structure, lithium diffusivity, and mechanical strength [179]. However, these properties can be quite different when they serve as the SEI on the surface of Li metal due to the complicated interfacial effects. Liu *et al.* investigated the interface between Li metal and two major SEI components, LiF and $Li_2CO_3$ [180]. The $Li_2CO_3$/Li interface has higher interfacial mechanical strength while the LiF/Li interface has higher electron tunneling energy barrier from Li metal to SEI. Accordingly, the contact between $Li_2CO_3$ and LiF can promote space charge accumulation along their interfaces, which generates a high ionic carrier concentration, significantly improves $Li^+$ ions transport and reduces electron leakage [181].

Many other simulations also investigated the interface system from various aspects, including the dimension effect, voltage effect, and interfacial polarization. For example, Pan *et al.* developed an informed space charge model to design an artificial SEI, which consists of LiF and $Li_2CO_3$ [182]. It was found that reducing the grain dimension of $Li_2CO_3$ in SEI can promote ionic carriers and increase the ionic conductivity by orders of magnitude. The effect of voltage is investigated by Leung and co-workers [183]. The applied voltage is manifested as changes in the electric double layer at the atomic scale, including charge separation and interfacial dipole moments. It was thus proposed that the manipulation on surface dipoles is a viable strategy to improve electrode passivation. Panahian Jand *et al.* studied the structure and mechanism of the growth of LiF on graphene [184]. The atomic structure of the SEI in $Li^+$ ion batteries is controlled by electrostatic effects and the stabilities of SEI can be predicted with a simple electrostatic model. Zhukovskii *et al.* performed comparative DFT-LCAO calculations on the polar interfaces inserted with extra Li atoms in the 2D interface [185]. The extra storage capacity at low potentials was also confirmed by this interfacial model. Moreover, Simeone *et al.* rationally designed a double-layer SEI for the first time [186]. This work provided a deep insight into the gold–electrolyte interface at the atomic scale based on the combination of experiment and theory.

Recently, artificial SEI is applied to resist the interfacial reactions, suppress the growth of lithium dendrites, and protect Li metal anode. A high-performance artificial SEI is supposed to meet several requirements. (1) It is stable against most components in electrolyte; (2) $Li^+$ ions can rapidly diffuse through it; (3) It possesses an enough high mechanical strength to suppress the growth of lithium dendrites; (4) Its stiffness is high enough to accommodate the deposition of Li metal during charging. A comprehensive theoretical study is very necessary to screen materials as artificial SEI due to the multifaceted requirements. Particularly, the proximity effect on artificial SEI induced by metal is very important due the close contact between Li metal and artificial SEI.

Materials with high stability and mechanical strength have been considered as artificial SEI. For example, a variety of layered materials (graphene, h-BN, *etc.*) with different topological structures were investigated as the artificial SEI for Li metal anode based on first-principles calculations by Tian *et al.* [44]. It is found that the defect type, the crystal structure, the dimension of the ring, and the appearance of metal proximity all make a great influence on determining the performance of artificial SEI, including the stability, the Li$^+$ ion diffusivity, and the mechanical strength. Specially, the introduction of defects can reduce the Li$^+$ ion transport impedance and promote the diffusion of Li$^+$ ions through the SEI layer yet greatly reduce the mechanical strength, lowering the stiffness, and critical strain and stress (Figure 13). The presence of defects can significantly weaken the mechanical strength of h-BN under equivalent biaxial tensile load and the emergence of metal can greatly reduce the Li$^+$ ion diffusion barrier. Further analyses confirmed the influence of the electrons in artificial SEI and the distribution of the charge on Li$^+$ ion diffusion barrier. The electron capture of Li$^+$ ion from SEI directly determines the diffusion barrier of Li$^+$ ion. The lower the charge density around the defect, the smaller resistance of the Li$^+$ ion diffusion. Simultaneously, the electrons in Li metal can transfer to the protective film, fill in the anti-binding orbital, and weaken the covalent bond in artificial SEI, which is the origination of the metal proximity effect. This study provides new insights into the interaction mechanism between Li$^+$ ion and artificial SEI materials at the atomic scale [44].

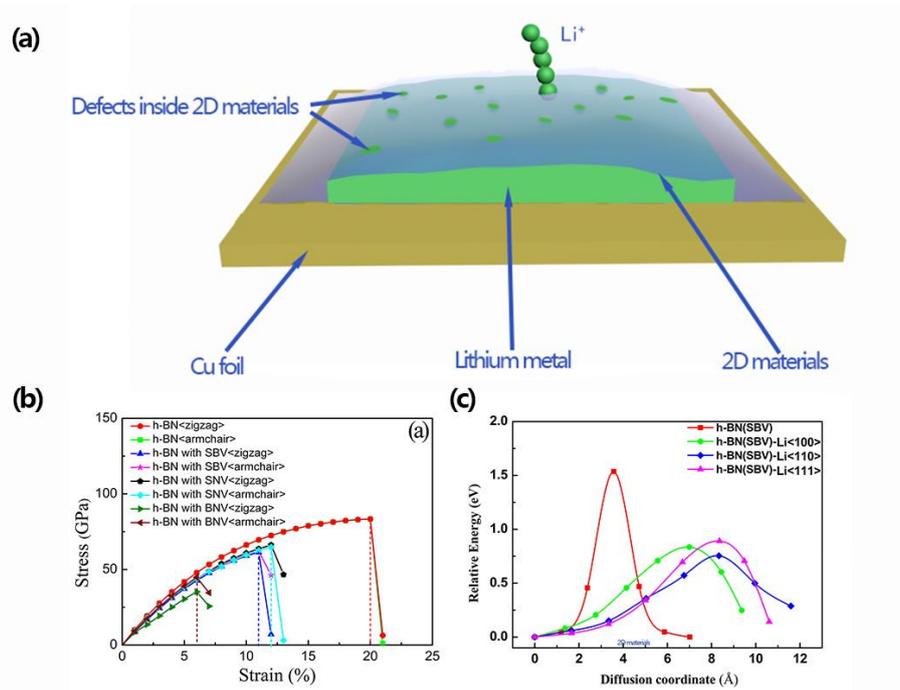

**Figure 13. (a)The diagram of Li metal anode structure; (b) The strain-stress curve of h-BN under equivalent biaxial tensile load; (c) The ion diffusion barrier through h-BN with different Li metal orientation [44].**

## 4.3. Lithium nucleation

Dendritic deposition is a common occurrence at high current for metal plating such as copper, nickel, zinc, and other metals. This problem is extremely prominent for active Li metal. In LMBs, there is a cation concentration gradient in the electrolyte between the two electrodes during electroplating. When the critical current density is reached, the current can only last for a short time before the cations are consumed in the electrolyte, which will disrupt the electrical neutrality on the surface of the plated electrode and create the localized space charge. As results, branched metal deposition will form. According to this theory, critical current density J* can be used to predict the dendrite growth. However, batteries typically operates at a current much less than J* in reality yet the dendritic can also happen, indicating that multiple mechanisms are working at the same time [34]. In this part, the key factors affect the morphology of Li deposition and the kinetic nucleation process are discussed.

In order to control Li metal anode surface morphology, researchers tried to study every parts of Li batteries to suppress the growth of Li dendrites and make stable SEI films. Recently, LMBs' cycle life can be substantially extended by optimizing the pulse charge current. Qi *et al.* [187] explored the mechanism involved in this process through molecular simulations. The pulsed charge frequency and amplitude can effectively control the solvation structure of $Li^+$ ions, and then significantly affect the diffusion of $Li^+$ ions (Figure 14b–c). Further experimental observation confirmed that the pulsed current can effectively control the solvation structure and diffusion process of $Li^+$ ions in the electrolyte, which can effectively suppress the dendrite growth (Figure 14a). They proved the interaction between anions and cations can improve the diffusivity of $Li^+$. The cycling life when using certain pulse current waveforms can be more than doubled. This research deepens the mechanistic understanding of the pulse charge process and provides a new perspective for the study of lithium dendrite growth.

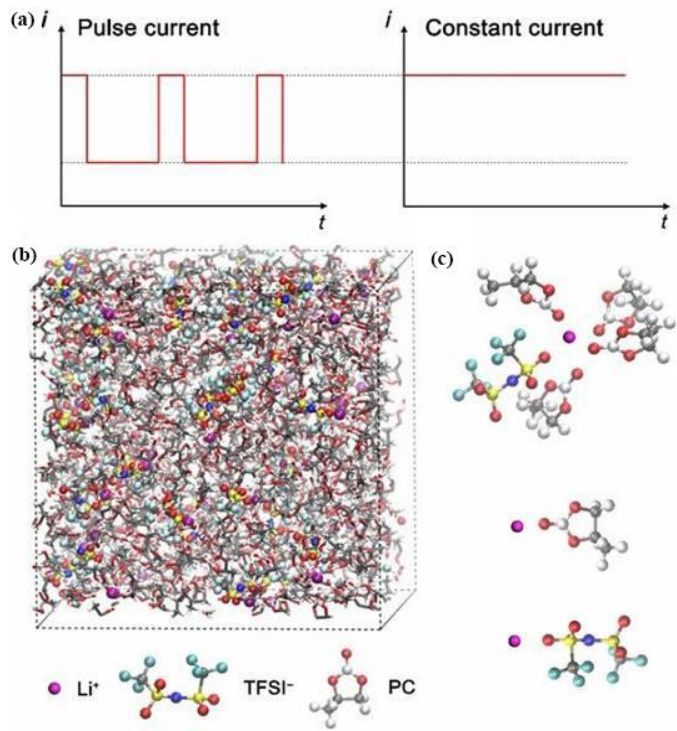

**Figure 14. MD simulations of 1 M LiTFSI in PC solvent. (a) Basic waveforms for pulse current and constant current. (b) Snapshot of the simulation box from MD simulations. (c) Representative configuration of solvation structures of Li$^+$ from MD simulations [187].**

Similar with Li–S and Li–O battery cathode, carbon materials can be also used for guiding the lithium nucleation. Considering the relatively weak affinity toward lithium by pristine material, nitrogen (N) doped-graphene was adopted as the lithium plating matrix to regulate Li metal nucleation and suppress dendrite growth by Zhang and co-authors [151]. The N-containing functional groups, such as pyridinic and pyrrolic nitrogen in N-doped graphene, are lithiophilic to guide lithium nucleation and thus deliver a uniform deposition. As a result, the N-doped graphene modified Li metal anode exhibits a dendrite-free morphology during repeated lithium plating and demonstrates a high Coulombic efficiency.

Beyond 2D carbons, nanodiamonds were used to regulate the lithium plating behavior and stabilize Li metal anode through a co-deposition with Li$^+$ ions [188]. Nanodiamonds can provide a more lithiophilic site for the deposition of lithium and a low diffusion energy barrier for uniform growth of lithium than routine copper current collector, delivering a stable cycling of lithium | lithium symmetrical cells. The first-principles calculation unveiled this phenomenon clearly. Nanodiamond (110) and Cu (111) surfaces with the lowest specific surface energies are selected to model lithium adsorption and diffusion (Figure 15a). Specifically, nanodiamnods possess a large binding energy of 3.51 eV toward Li$^+$ ions due to a large amount of electronic migration, which is around 1 eV higher than that of copper (Figure 15b). Besides, the lowest diffusion barrier of Li$^+$ ion on nanodiamnod surface among considered species, including various SEI components ($Li_2O$, $Li_2CO_3$, LiOH, LiF, *etc.*) and Li metal, induces a quick diffusion and a uniform deposition to deliver a dendrite-free morphology (Figure 15c–d).

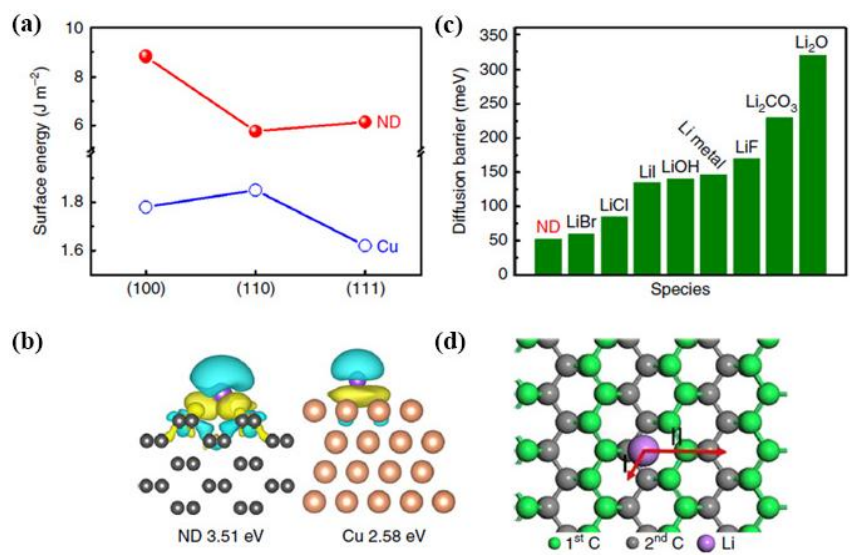

**Figure 15. First-principles calculations to describe $Li^+$ ion plating behavior on nanodiamond surface. (a) Surface energies of low index facets for nanodiamond and Cu. (b) Differences of charge density for Li on nanodiamond (110) and Cu (111) surfaces. (c) Diffusion barrier of Li on different surfaces. (d) The most stable adsorption sites and diffusion paths for Li on nanodiamond (110) surface [188].**

Up to now, theoretical simulations have achieved a great progress in Li metal anode researches, affording new insights into the understanding on the modification of SEI, the construction of a composite anodes, and the complicated interfacial reactions between electrolytes and anodes. Despite the great achievements, some fundamental and important issues remain unclear due to various challenges. The development of new techniques in combination with first-principles calculations is supposed to provide a new chance to solve these intractable issues. For example, recently, cryo-electron microscopy revealed the detailed lithiation growth directly and provided a complete understanding of the mechanism of batteries failure issues [189]. Theoretical simulation and cryo-electron microscopy can be combined to illustrate the mechanism of lithium nucleation. It is thus strongly believed that theoretical simulations can play a more and more important role in investigating the very complicated interfacial reactions with more reliable models as the development of computer technology, theoretical chemistry, and experimental characterization techniques.

# 5. Modeling of SSEs

Solid-state electrolytes is one of the key techniques to break through the energy density bottleneck of conventional lithium batteries. Besides, it is also a research hotspot in the field of electrochemical energy storage [190, 191]. Because the SSEs are independence of the liquid organic solvents, there are not leakage issue and flatulence burning problem in solid-state battery. The SSEs also fundamentally solve the problems of inadequate chemical and electrochemical stabilities, low ion selectivity, and poor safety in conventional LIBs [192-194]. In addition, SSE exhibits a larger mechanical strength, which can also efficiently suppress the dendritic growth of lithium. Up to now, various kinds of the mineral structured lithium compound have been demonstrated to be excellent SSEs (Figure 16a–d). However, comparing with liquid electrolyte-based batteries, the charge and discharge speed in solid-state batteries is much slower due to the much slower lithium diffusion inside the solid material than that in liquid electrolytes. On the other hand, the contact between SSEs and anodes is poor due to the rigidity of the SSE, which leads to limited active sites and a large interfacial resistance. Therefore, it is very crucial to design solid-state electrolytes with super lithium ion conductivity, high chemical and electrochemical stability, and a low interfacial impendence with electrodes. As concerned, investigating the Li$^+$ ion diffusion mechanism in solid-state electrolytes is a very important topic [20, 195, 196].

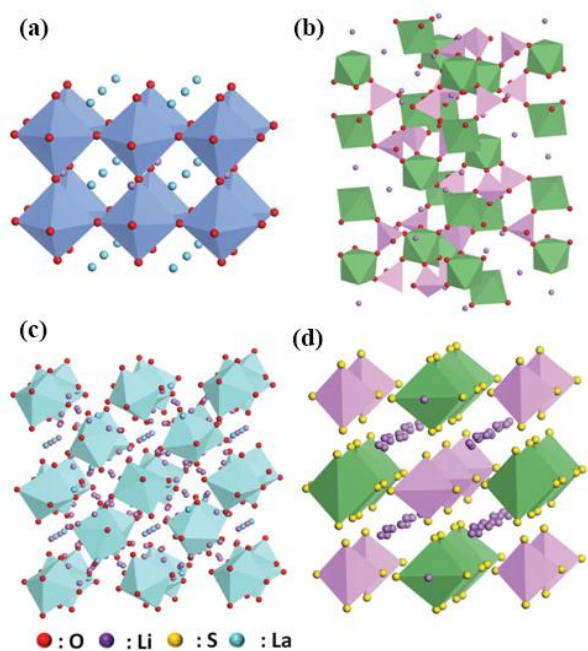

**Figure 16. Typical solid-state electrolytes. (a) Perovskite-type LLTO. (b) NASICON-type LAGP. (c) Garnet-type LLZO. and (d) thio-LISICON-type LGPS [196].**

A lot of progresses have been made in solid-state electrolytes through theoretical simulations. Firstly, the modeling of solid-state materials and solid-based electronic interaction can be free from complicated liquid environment, which is relative easier to carry out. Secondly, due to the variety of the multicomponent solid-state electrolytes, first-principles calculation has the overwhelming advantage on efficiently searching for the stable structures and achieve the key parameters that determines their performance as intermediate electrolyte. Two major issues should be concerned theoretically: one is the exploration of the solid-state electrolyte material with chemical and electrochemical stability as well as excellent lithium ion conductivity; the other is the optimization of interfacial impedance and interfacial stability. Many theoretical schemes have been developed to optimize the property and efficiency of SSEs [197]. Among them, most encouraging progress has been made through high-throughput screening based on material genome method.

## 5.1. Screening of super Li$^+$ conductors

High-performance SSEs should simultaneously satisfy several requirements: electronic insulating, high Li$^+$ ion conductivity, and high chemical and electrochemical stability. All of these properties can be simulated by first-principles calculations. In 2011, Kanno *et al.* [198, 199] reported a new three-dimensional structure solid-state electrolyte, $Li_{10}GeP_2S_{12}$ (LGPS). The ionic conductivity can be as high as 12 mS cm$^{-1}$ at room temperature, which is much higher than that of traditional SEEs, such as $Li_2S$-$GeS_2$, $Li_2S$-$P_2S_5$, $Li_2S$-$B_2S_3$, and $Li_2S$-$SiS_2$ (conductivity higher than $10^{-4}$ S cm$^{-1}$ at room temperature). Specially, Mo *et al.* [200] systematically simulated various properties of LPGS, including the activation barrier (Li diffusivity), the chemical stability, and the electrochemical window. According to the computation, LGPS possesses a high lithium conductivity of 12 mS cm$^{-1}$ at room temperature, a low activation energy of approximate 0.1 eV, a high lithium transference of 0.99, and a stability electrochemical window from 0 to 6 V *vs.* Li/Li$^+$. This work not only perfectly clarified the excellent performance of $Li_{10}GeP_2S_{12}$ in experiment, but also opened up the opportunity for the theoretical prediction and optimization of lithium super ionic conductor material.

Beyond LGPS, the Li$^+$ conductivity and chemical stability of many other SSEs have been investigated through first-principles calculations. Ong *et al.* [201] investigated the phase stability, electrochemical stability and Li$^+$ conductivity of the $Li_{10\pm1}MP_2X_{12}$ (M = Ge, Si, Sn, Al or P, and X = O, S or Se) family of super ionic conductors. It was found that the $Li_{10}GeP_2S_{12}$ has the highest Li$^+$ conductivity among this series of materials. The results also demonstrated that isovalent cation substitutions of Ge$^{4+}$ have a small effect on relevant intrinsic properties. Therefore, $Li_{10}SiP_2S_{12}$ and $Li_{10}SnP_2S_{12}$ possess similar phase stability, electrochemical stability, and Li$^+$ conductivity as that of LGPS. Then, the similar method was applied to various SSEs [36], such as $Li_7La_3Zr_2O_{12}$ (LLZO), $Li_{6.5}La_{3-x}Ba_xZr_{1.5-x}Ta_{0.5+x}O_{12}$(LLBZTO) [202], and a flexible composite solid-state electrolyte membrane consisting of inorganic solid particles ($Li_{1.3}Al_{0.3}Ti_{1.7}(PO_4)_3$), polyethylene oxide (PEO), and boronized polyethylene glycol (BPEG) [12].

Although all of the properties can be theoretically predicted by simulations, a large-scale computation is necessary for an efficient exploration on excellent SSEs due to the great diversity of potential solid-state electrolytes and various modification strategies. The recent developed high-throughput simulation approach is viewed as an ideal scheme for the screening of SSEs [203]. Sendek *et al.* [24] presented a new type of computational approach to identify promising candidates for SSEs, which is capable of screening all known lithium containing solids. Following the procedure shown in Figure 16, they firstly screened 12831 lithium containing crystalline solids with high structural and chemical stability, low electronic conductivity, and low cost. Then they developed a data-driven ionic conductivity classification model using logistic regression for identifying the candidate structures that are likely to exhibit fast lithium conductivity based on reported experiments. The screening program reduces the list of candidate materials from 12831 to 21 structures that show promising features as electrolytes, few of which have also been examined experimentally. Although none of simple atomistic descriptor functions alone provide reasonable prediction for ionic conductivity, a multi-descriptor model can be useful. The screening for structural stability, chemical stability, and low electronic conductivity eliminates 92.2% of all Li-containing materials while the screening for high ionic conductivity eliminates a further 93.3% of the remainders.

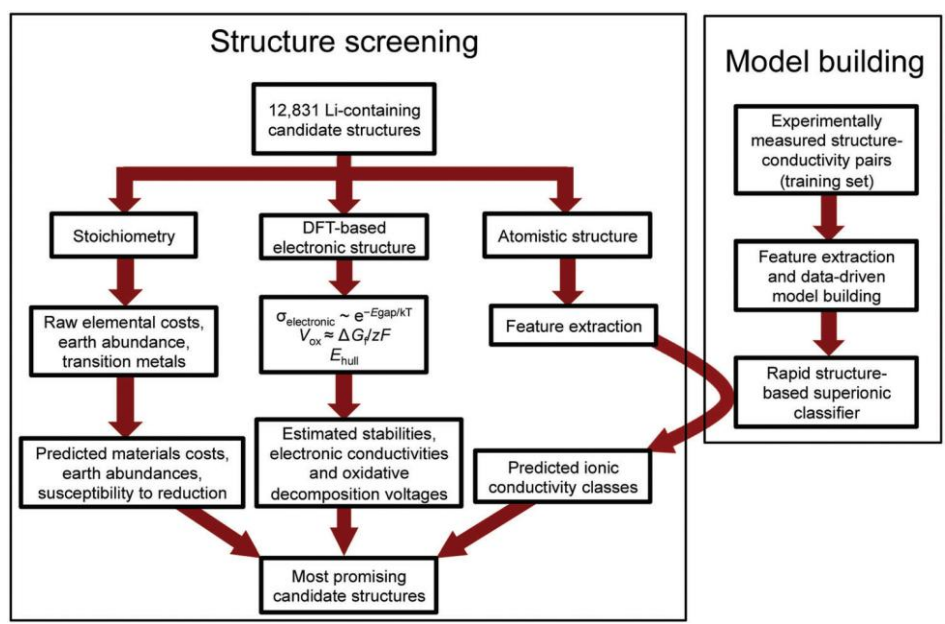

**Figure 17. Flowchart of the approach for screening high-performance solid-state electrolytes [24].**

For the application in battery, solid-state electrolytes should match well with cathodes, which is one of the more critical considerations beyond its intrinsic properties. In order to find a solid-state electrolyte match well with cathodes, Richards *et al.* [204] developed a high-throughput computational methodology, integrating the material database and DFT computation. It was found that thiophosphate electrolytes have especially high reactivity toward high voltage cathodes and a narrow electrochemical stability window. This suggested a few combinations of known cathode and electrolyte materials can deliver a high-performance battery, such as the combination of $Li_3PS_4$ or $Li_7P_3S_{11}$ and $LiVS_2$. They also find that a number of known electrolytes are not inherently stable yet react with the electrode to form passivating but ionically conducting barrier layers, such as LiPON. Electrolytes with high chemical and electrochemical stability can be screened through this efficient method, which simultaneously reduces experimental expense and save time.

The high-throughput scheme main focused on the phases has been experimentally identified and already recorded in database. By applying the structure prediction computation, new SEEs can be theoretically predicted, which opens up a new opportunity of designing novel SEEs. Recently, Wang *et al*. [205] designed a novel oxygen-sulfide solid electrolyte, LiAlSO, of which the crystal structure was confirmed by high-throughput calculation based on popular crystal structure prediction method. The thermodynamic stability, kinetic stability, and ion transport properties were also studied. Through the development of barrier computing program BVpath, based on semi-empirical potential, the researchers combined different methods of calculation accuracy for different stages of material screening and optimization. The calculated activation pathway (Figure 18a) and barrier calculation (Figure 18b–d) showed that LiAlSO has a very low energy barrier for $Li^+$ ions migration along the *a*-axis direction (PATH I, as small as 0.0064 eV). This is the first type of SSE developed by the idea of structure-prediction based material genomics. Moreover, solid-state electrolytes were extended to oxysulfides and mixed anionic compounds.

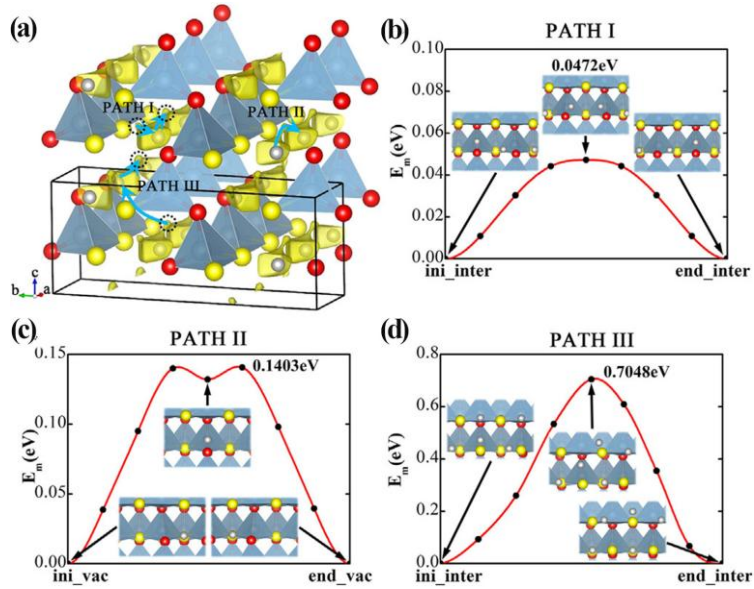

**Figure 18.** The calculated kinetic properties of LiAlSO. (a) The continuous Li$^+$ migration channels determined by the BV-based method, and the energy barriers simulated by the NEB method. (b) The kick-off mechanism along the direction. (c) The direct vacancy hopping along the *a*-axis. (d) The interstitial Li$^+$ hopping along the *c*-axis [205].

## 5.2. Li$^+$ ion migration mechanism

Although solid-state electrolytes can address the flammability and limited electrochemical stability, the Li$^+$ ion conductivity in SSEs is much lower than that in liquid. The Li concentration, the crystalline structures, and the doping element can all make great influence on the ion conductivity. Therefore, a deep understanding of the ion diffusion mechanism is quite important. First-principles calculations can provide a fruitful insight into the ionic transport mechanism and guide the further discovery of new SSEs. In 2010, Li *et al.* [206] employed first-principle method to explore the diffusion mechanism of Li$^+$ ion in α- and β-Li$_3$N. They found the diffusion of Li$^+$ ion in α- and β-Li$_3$N occurs in different planes. At the same time, Li$^+$ ion migration energy barriers were also calculated.

Based on the systematic research of the diffusivity in SSEs, Wang *et al.* [35] revealed a fundamental relationship between anion packing and ionic transport in fast Li-conducting materials by investigating various solid sulfide materials. Through a comparative activation barrier simulation in the bcc/fcc/hcp lattices at different volumes (Figure 19a–c), it is found that an underlying body-centered cubic-like anion framework, which allows direct Li hops between adjacent tetrahedral sites, is the most desirable topological structure for achieving high ionic conductivity. Such type of anion arrangement is indeed present in several known fast

Li-conducting materials and other fast ion conductors. These findings provide an important insight toward lithium ion transport in SSEs and serve as an important principle for further discovery and design of high-performance SSEs.

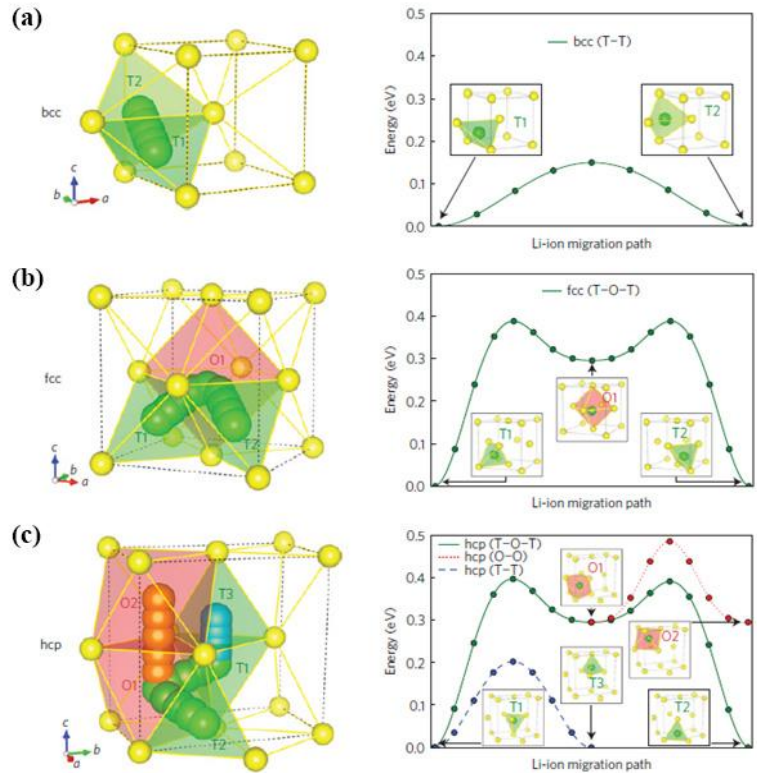

**Figure 19. Li-ion migration pathways in different type of anion lattices. (a) Li-ion migration path (left) and calculated energy path (right) in bcc sulphur lattices. (b) Li-ion migration path (left) and calculated energy path (right) in fcc sulphur lattices. (c) Li-ion migration path (left) and calculated energy path (right) in hcp sulphur lattices. The sulphur anions are colored yellow, and the Li$^+$ ions are colored green, blue, and red for different paths, respectively. LiS$_4$ tetrahedral and LiS$_6$ octahedral are colored green and red, respectively [35].**

Through theoretical researches, Fang *et al.* [207] found the Li$^+$ ions superconductivity can be achieved by using cluster ions instead of basic ions based on Li$_3$SBF$_4$ or Li$_3$S(BF$_4$)$_{0.5}$Cl$_{0.5}$ solid-state electrolytes. The cluster ions-based SSEs not only exhibited a very high ion conductivity (>10$^{-2}$ s/cm) at room temperature, but also had the advantages of a large band gap and high melting point. The comparison between the two kinds of ions revealed different characteristics of cluster ions from those of elementary ions, such as ultrahigh electron affinity, large dimension, and internal charge fractions. The vibration mode and dimension effect induced by cluster ion also promote the migration of Li$^+$ ions in SSEs. Furthermore, the use of cluster ions can be served as a basis to construct an excellent SSE with ultrahigh electron affinity, large dimension, internal charge fractions, and thus high ion conductivity. The research theoretically demonstrated the usage of cluster atoms instead of elementary atoms to form bulk energy materials and revealed the changes to the bulk material properties due to the special physical and chemical properties of cluster ions.

It is commonly believed that the $Li^+$ ion transports one by one through the solid-state electrolytes and single-Li model dominates the theoretical research in various lithium battery system. Recently, however, He *et al.* [208] revealed a unique mechanism of fast-ion conductor material's "coordinated movement" and demonstrated multi-ion coordinated movement plays an important role in reducing the energy barrier of ion transition and the rapid transport of excited ions. The energy of each ion is different in all multi-ion coordinated transitions, and some ions are in the position with high energy while the others are in the position with low energy. When moving together, the ions in the high-energy position move downward to partially offset the energy barrier perceived by the upward movement in the low-energy position. Therefore, the cooperative movement of multiple ions exhibit a lower energy barrier comparing with the single ion movement (Figure 20). Based on this theoretical model, a new material design strategy to improve the lithium ion conductivity in solids can be achieved. By doping the material, part of the $Li^+$ ions were in the position with high energy position, and these ions can stimulate the cooperative transition, thereby transform the original ordinary material into a fast ionic conductor material. Such material design strategy has already been confirmed in several new fast ionic conductors. This work gain novel insight on the diffusion mechanism in solid-state electrolytes and provides a theoretical guidance for designing fast ionic conductor materials.

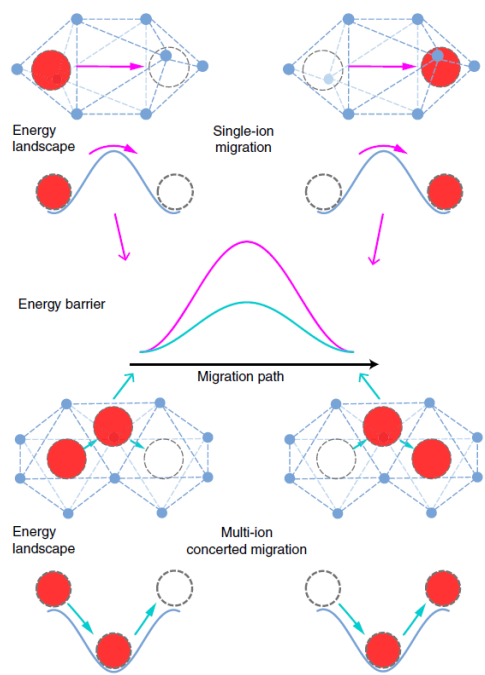

**Figure 20.** Schematic illustration of the comparison between single-ion migration and multi-ion concerted migration. The simulation results exhibited that the multi-ion migration induced much lower barrier [208].

## 5.3. Interphase between SSEs and electrodes

Optimizing the interfacial stability and interfacial impendence is another major challenge hindering the large scale application of SSEs. Therefore, the interfacial effect between two solid materials should be clarified in advance. Such challenge is especially remarkable for the active Li metal anode [156, 209, 210]. On one hand, SSEs should be chemically and physically stable at the interface region. On the other hand, the diffusion of Li$^+$ ion cannot experience too high energy barrier to promise the high interfacial conductivity. As results, the research on the interfacial properties becomes quite necessary. Such interfacial conductance is especially dramatic for the garnet-type SSE, although they possess high ionic conductivity, excellent environmental stability, and wide electrochemical stability window [195].

The reaction mechanism between SSE and electrode is crucial to understand the interfacial geometry, interfacial stability, and interfacial properties. Zhu *et al.* [211]found that the outstanding stability of SSEs is mainly originated from kinetic stabilizations. The sluggish kinetics of the decomposition reactions causes a high overpotential leading to a nominally wide electrochemical window. The decomposition products can gradually form on the interface, which mitigates the extreme chemical potential from electrodes and protect solid-state electrolyte from further decompositions. They further revealed the passivation mechanism of these decomposition interphases and quantified the extensions of the electrochemical window from the interphases. A new understanding was thus achieved on the general principles for designing SSEs with enhanced stability and engineering interfaces. In another study, they identifies the chemical and electrochemical stabilities of the interphase further from thermodynamic aspect. It is found that many solid electrolyte–electrode interphases have limited chemical and electrochemical stability. Although the formation of interphase layers is thermodynamically favorable at these interfaces, the formed interphase exhibits different properties [212]. Miara *et al.* [213] also simulated the decomposition reactions with high fidelity, indicating that dense cathode composites with high impedance can be formed between spinel cathodes and oxide electrolyte.

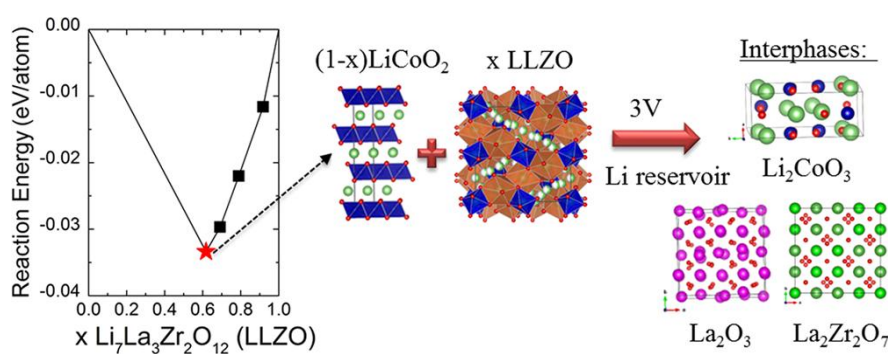

**Figure 21. Interfacial products produced by spinel cathodes and oxide electrolytes.**

Another scheme to optimize the interfacial properties is to modify SSEs by introducing defects or dopants. Miara *et al.* calculated the defect energies and site preference of all possible dopants in $Li_7La_3Zr_2O_{12}$ (LLZO), which is viewed as an ideal candidate for all-solid-state lithium batteries. The findings suggested that several novel dopants, such as $Zn^{2+}$ and $Mg^{2+}$, can be stable on the Li-sites and Zr-sites (Figure 21). To understand the source of interfacial resistance between the electrolyte and cathode, they investigate the thermodynamic stability of the electrolyte–cathode interphase by calculating the reaction energy for LLMO (M = Zr, Ta) against $LiCoO_2$, $LiMnO_2$, and $LiFePO_4$ cathodes over the voltage range seen in lithium-ion battery operation. These results provide a reference for use by researchers interested in bonding these electrolytes to cathodes.

Mechanical properties of the interphase between solid-state electrolyte and anode is also important. This especially important for Li metal anode due to the requirements of resisting lithium dendrites. Monroe *et al.* [214] proposed the "double shear modulus theory" to predict whether the electrolyte can inhibit dendritic growth. Specifically, when the electrolyte shear modulus is greater than 2 times of the shear modulus of metallic lithium, the interface can be viewed as stable. Interfacial roughening is mechanically suppressed when the separator shear modulus is about twice that of lithium. However, the mechanical properties of most Li-ion super conductor is not good. Lu *et al.* [215] used quantum chemistry calculation to systematically investigate a series of metal boron hydrides ($Li_2B_{12}H_{12}$, $LiCB_{11}H_{12}$, *etc.*) with high ionic conductivity [216, 217]. It was found that the shear modulus is relatively small. Therefore, this solid-state electrolyte is not likely to inhibit the dendritic growth of lithium, although they exhibited chemical stability toward Li metal. The mutual restriction between Li conductivity and mechanical strength is one of the most difficult points for designing electrolyte and protective layer. When the ion conductivity is high, the electronic interaction between $Li^+$ ion and surrounding materials should be weak enough for $Li^+$ ion to escape from the impedance imposed by surrounding ions, and the host material should offer a large space for the diffusion. However, such a weak interaction or large space should always induce a low strength for host materials. This contradiction is also clarified in the work on the protective film of Li metal anode by Tian *et al.* [44].

Recently, the artificial plating of ultrathin intermediate layer on the interfacial region is found to be an effective way to solve the poor conductance and suitable types of the intermediate materials have been theoretically investigated. Han *et al.* [218] effectively addressed the large interfacial impedance between a Li metal anode and garnet electrolyte using ultrathin $Al_2O_3$ by atomic layer deposition. The computational results revealed that the oxide coating enables wetting of metallic lithium in contact with the garnet electrolyte surface and the lithiated-alumina interface allows effective $Li^+$ ions transport. Wei *et al.* [219] found that the chemical properties of a promising oxide-based SSE (garnet) changed from "super-lithiophobicity" to "super-lithiophilicity" through an ultrathin coating of Si. The first-principles calculations were performed to investigate the interfacial stability and the lithium diffusion between LLZO and the newly formed lithiated Si. In another work, this group proposed a new methodology to reduce the garnet/lithium-metal interfacial resistance by depositing a thin Ge layer on garnet SSE. The first-principles calculations also confirmed the good stability and improved wetting at the interface between the lithiated Ge layer and garnet. These series of works bring a new opportunity on designing a new coating layer between electrolyte and Li metal anode.

So far, only a few materials are known to be stable against Li metal, and people's interest mainly focused on few kinds of ordinary lithium-based material to form the interface, especially for the Li metal. To resolve this outstanding challenge, lithium-against-stable materials have been unearthing out of chemistry across the periodic table using high-throughput calculations based on large materials database. It is found that most oxides, sulfides, and halides, commonly studied as protection materials, are reduced by Li metal due to the reduction of metal cations. Nitride anion, in contrast, exhibits unique chemical stability against Li metal, which is either thermodynamically intrinsic or a result of stable passivation. In general, metal nitrides have a significantly lower reduction potential than oxides, sulfides, and fluorides. Many nitrides are thermodynamically stable against Li metal. On the basis of this chemistry knowledge, they suggest new strategies to form stable SEI on Li metal anode. Specifically, many cations, such as Mg, Ca, B, Al, Zr, V, W, Si, Ti, Nb, and Ta, can be protected in nitride anion chemistry systems at sufficiently high nitrogen content. However, some cations, such as Ge, Sn, Ga, and Zn, are always reduced by Li metal regardless of anion chemistry and composition. The results herein establish essential guidelines for selecting, designing, and discovering materials for Li metal protection, while nitride materials or high nitrogen doping is an alternative choice for building stable rechargeable LMBs [220].

Despite great theoretical progress on SSEs, there are still some key challenges for all-solid-state lithium batteries such as volume change of electrode , interfacial charge transfer resistance, inflexibility, and poor cycling stability [220]. In general, it is a useful endeavor to encourage the study of SSEs to build a practical solid-state LMBs. Not only intrinsic properties of SSEs but also the interfacial reactions between SSEs and anodes should be taken into consideration. Therefore, the exploration of new SSEs and the further investigation on the theoretical mechanism are still crucial topics, in which theoretical simulations can play an important role [221].

# Outlook

We have reviewed next-generation LMBs for a theoretical viewpoint, including sulfur cathodes, oxygen cathodes, lithium metal anodes, and solid-state electrolytes. Tremendous fruitful achievements have been obtained in terms of explaining experimental observations and delivering a rational design strategy of various electrode or electrolyte materials. It has been widely recognized that theoretical simulation has become an indispensable technique in battery and material researches, which can provide a deep insight at the atomic level. However, there are still many issues remain unclear, which seriously hinder the further improvement of lithium metal batteries. The difficulty of theoretical researches lies on several aspects, and targeted research plan can be made accordingly:

1. It is still very hard to achieve a reasonable empirical potential or tight binding parameter for lithium ion. As results, a full-quantum description of lithium-containing system is requisite and first-principle calculations become almost the only choice for such simulations. However, the simulated system based on first-principle calculations is seriously limited. Simulations of meso- or micro-dimension systems, which is commonly required in LMBs and can provide more information than simulations of smaller systems, is almost impossible. Although empirical methods have been used in some works, potential functional does not exhibit the universality of rationally describing different battery systems, especially for the complicated interface models. Therefore, developing empirical methods is the potential pathway to break through the limitation of the currently used method and greatly expands the modeling object. Among various methods, the newly developed empirical potential fitting based on neural network is a good choice to build a reasonable potential [222, 223].

2. The lithiation and delithiation processes in LMBs are quite different from that in traditional LIB with a relatively unchanged framework. The structure and phase of the active materials experience fundamental transformation as the lithium concentration changes in LMBs. Such transformation is neither easy to be identified in experiment nor predicted by traditional simulation scheme. Both the structure and properties of the intermediate phases are very hard to be identified. Such problem is extremely extrusive for Li–S batteries. The recently developed high-throughput method, which has already been applied in SSEs, is one of the most promising pathway to solve the current problems in electrode, search for stable structure and material systems, and predict the new phases during charge-discharge cycling. This method can provide a good instance for studying both thermodynamics and kinetics in LMBs. One the other hand, the high-throughput approach for low-dimensional system is also necessary to be further developed due to the importance of interfacial structure and two-dimensional materials in batteries.

3. The kinetic properties during lithiation/delithiation process is critical for building the electrochemical theory in next-generation LMBs. Such issue is especially important for the study of the interfacial region, such as the electrolyte–electrode, active-species–electrode-framework, and active-species–electrolyte interfaces. The study of the kinetic properties should be performed on the basis of the identification of the structure and the clarification of the thermodynamic properties. The progress in SSEs can inspire the researchers in other fields to carry out the kinetic properties simulation by applying similar methods. Simultaneously, new modeling scheme should be developed to satisfy the unique characteristics of different battery systems.

# Conclusion

In this review, we have summarized recent theoretical progress of next-generation lithium metal batteries, including sulfur cathodes, oxygen cathodes, lithium metal anodes, and solid-state electrolytes. With an introduction of the current research state, the major concerned theoretical problem, the representative works in various fields, the remaining problem, and the potential research directions in the future are summarized. Despite encouraging progresses, there are still many unsolved problem in this filed. New idea and method are all highly required for the further development of the modeling and computation. The application of first-principles calculations in LMBs also meets some limitations currently. Despite of the gap between theory and experiments due to the differences between complicated experimental condition and ideal equilibrium theoretical condition, theoretical results are still helpful to explaining experimental observations and delivering a rational design strategy of various electrode or electrolyte materials. The success of theoretical study on traditional LIB have demonstrated the significant role of computational design to achieve an excellent battery performance. Therefore, it is strongly believed that theoretical modeling has a bright future and can play a promising role in next-generation lithium metal batteries.

# Modeling of Next-Generation Lithium Metal Batteries

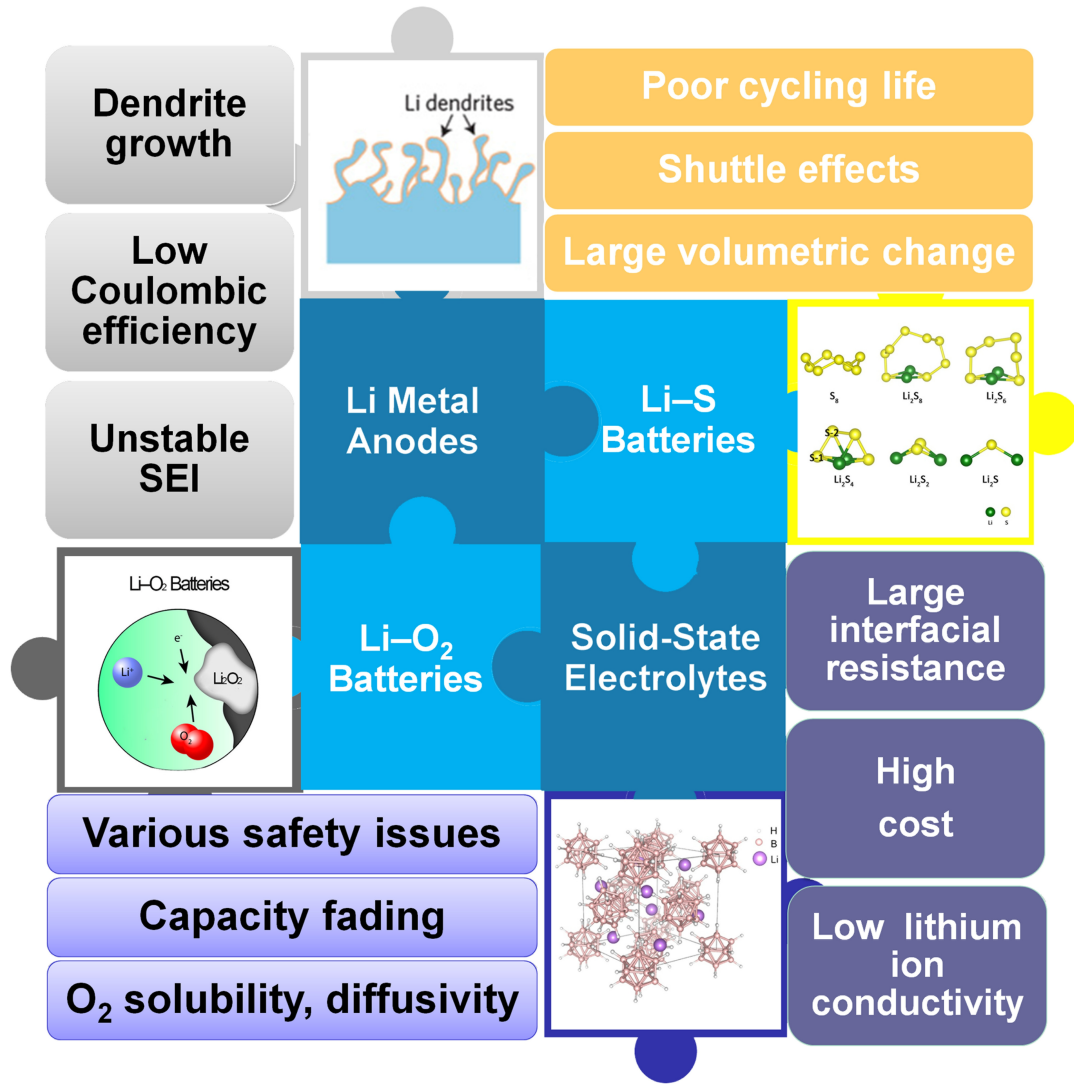

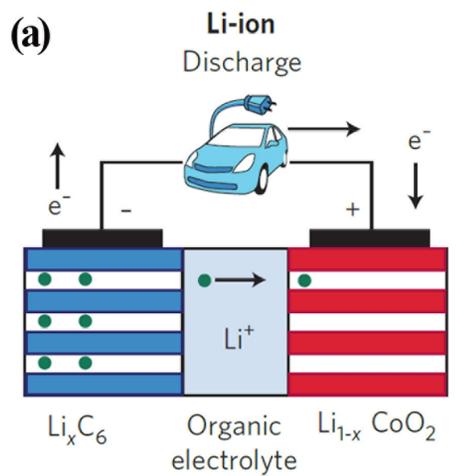
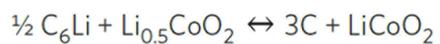
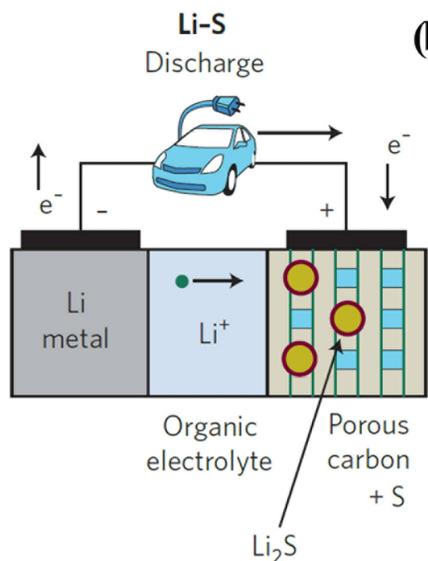
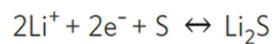
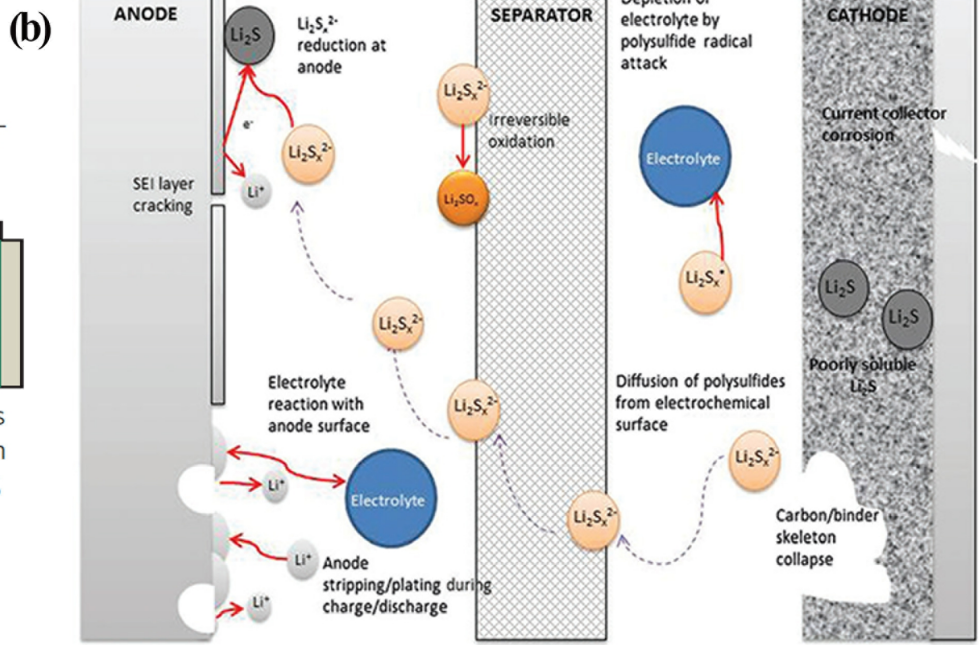

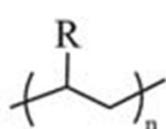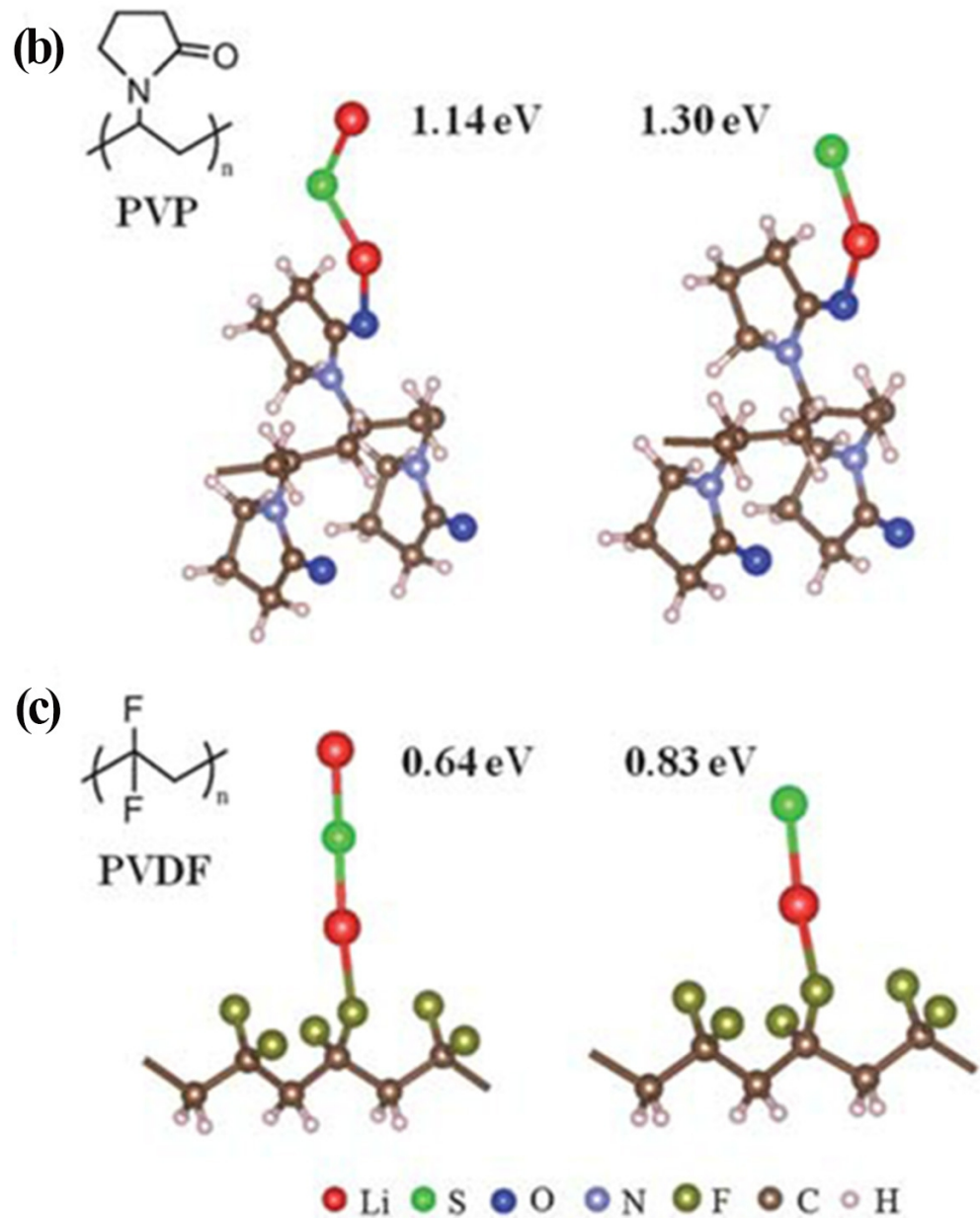

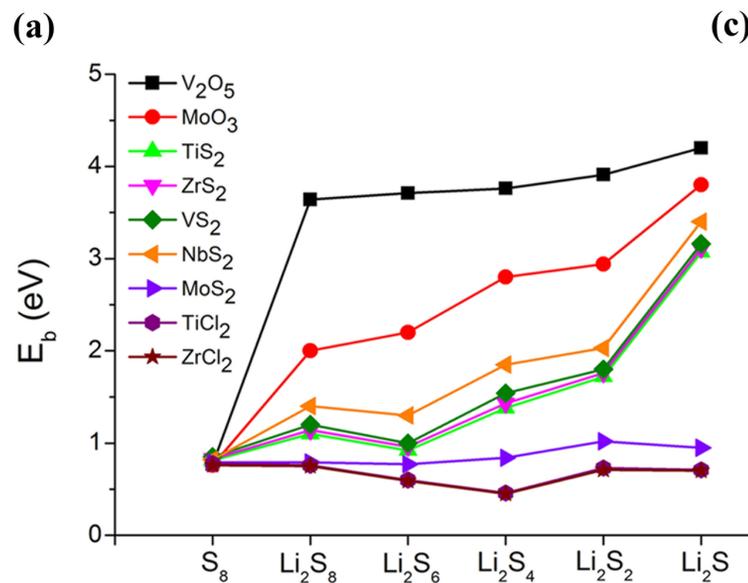
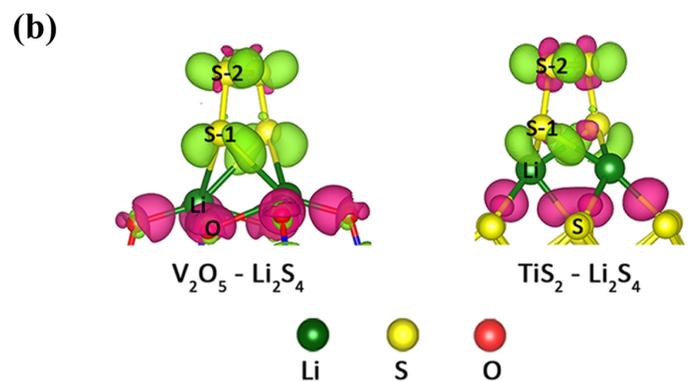
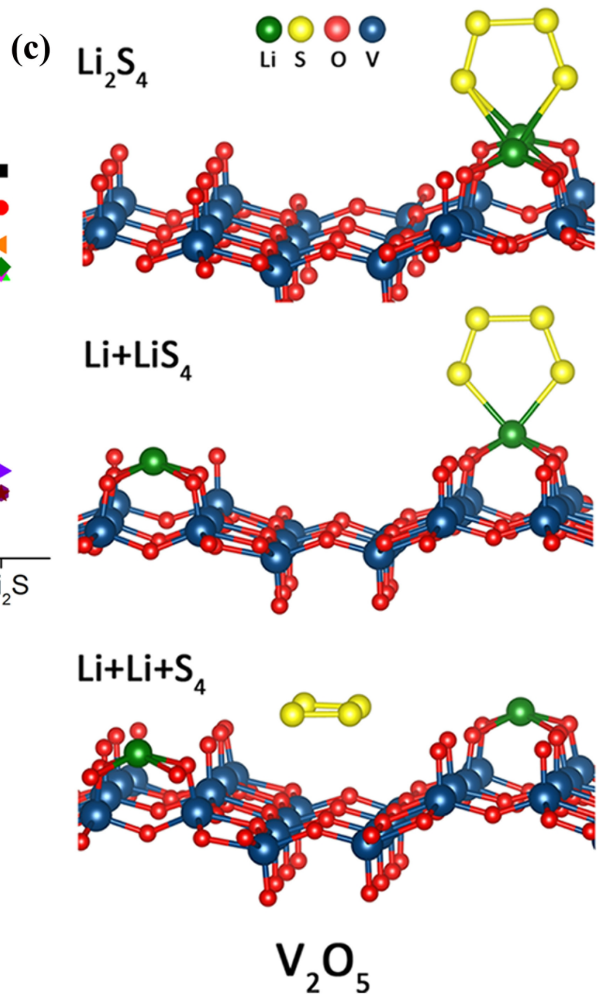

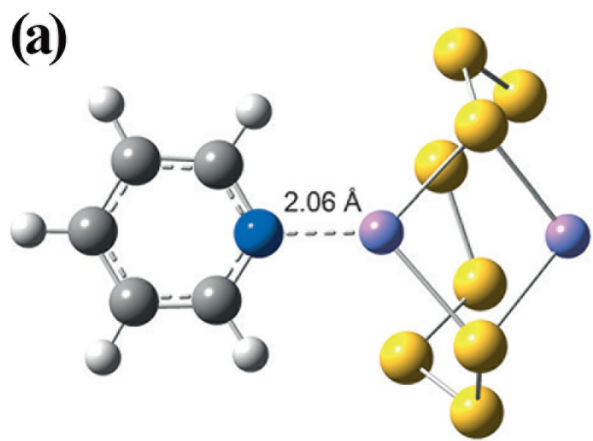
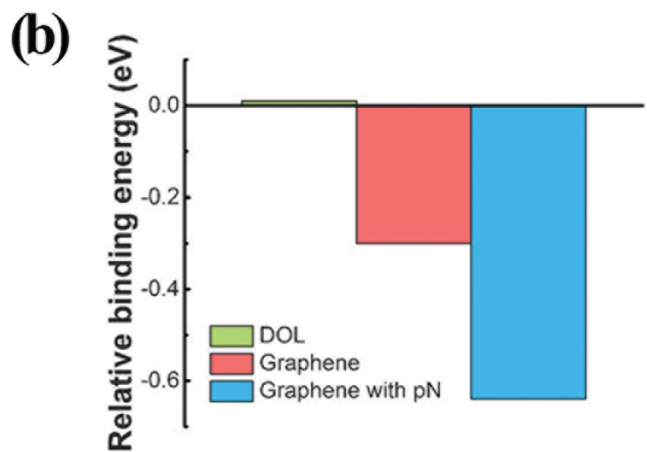
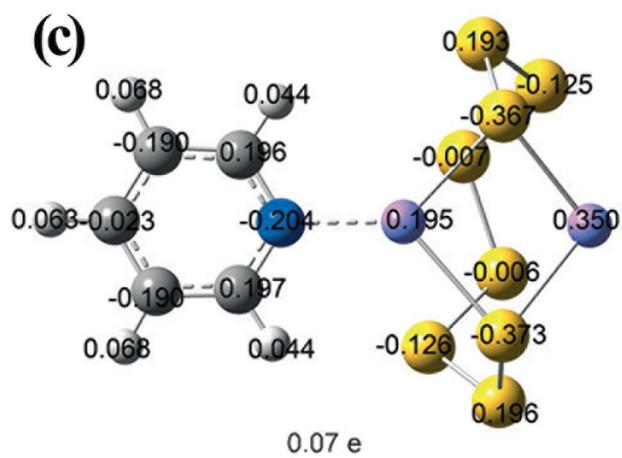
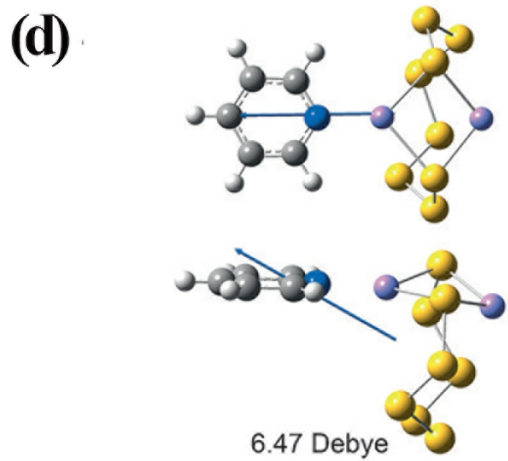
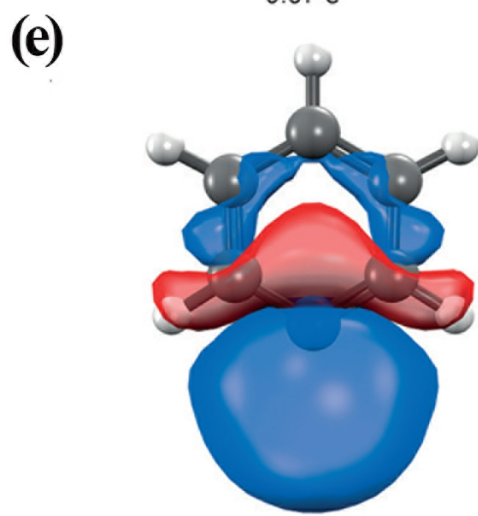
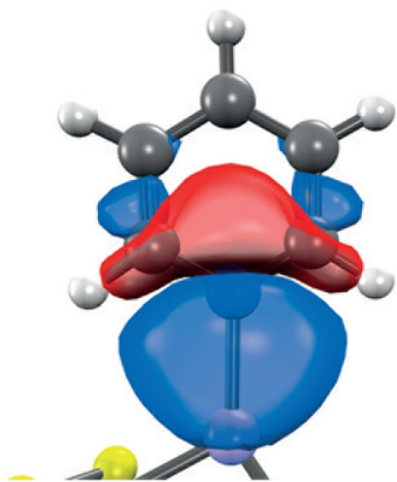

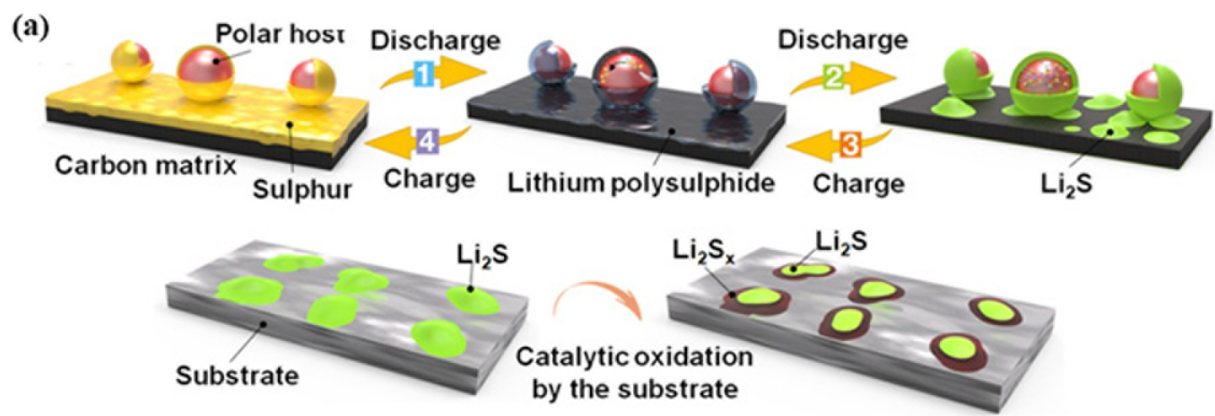
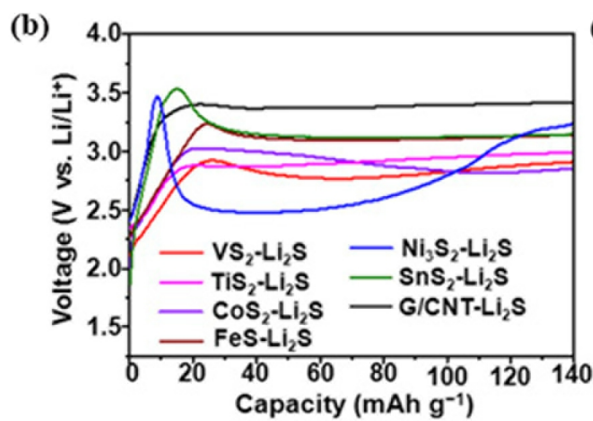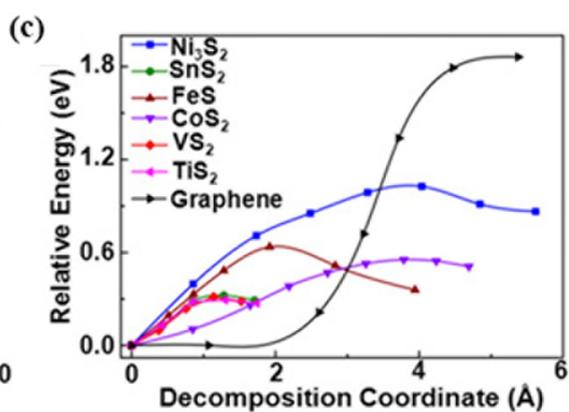
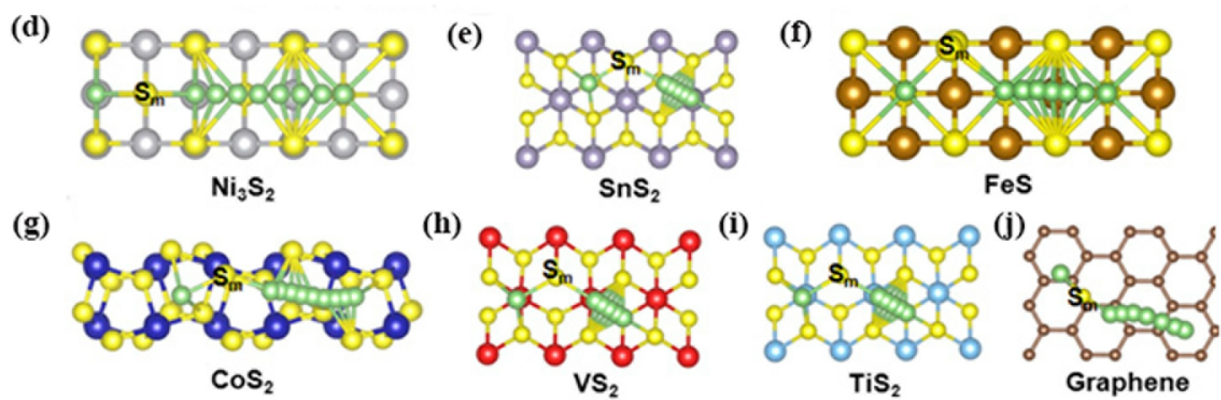

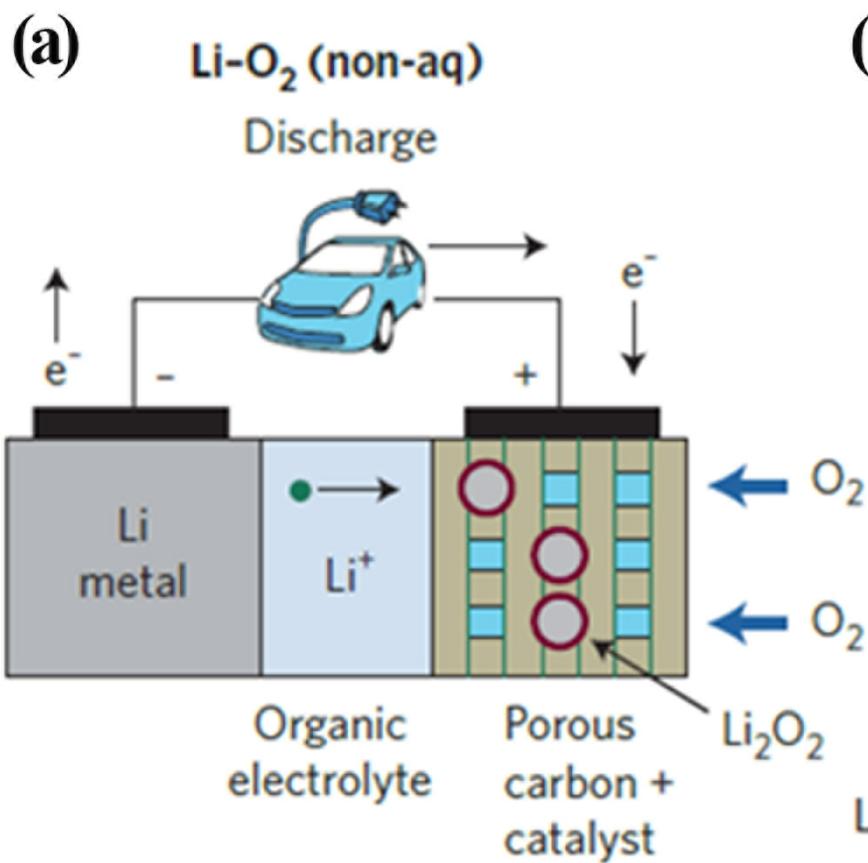

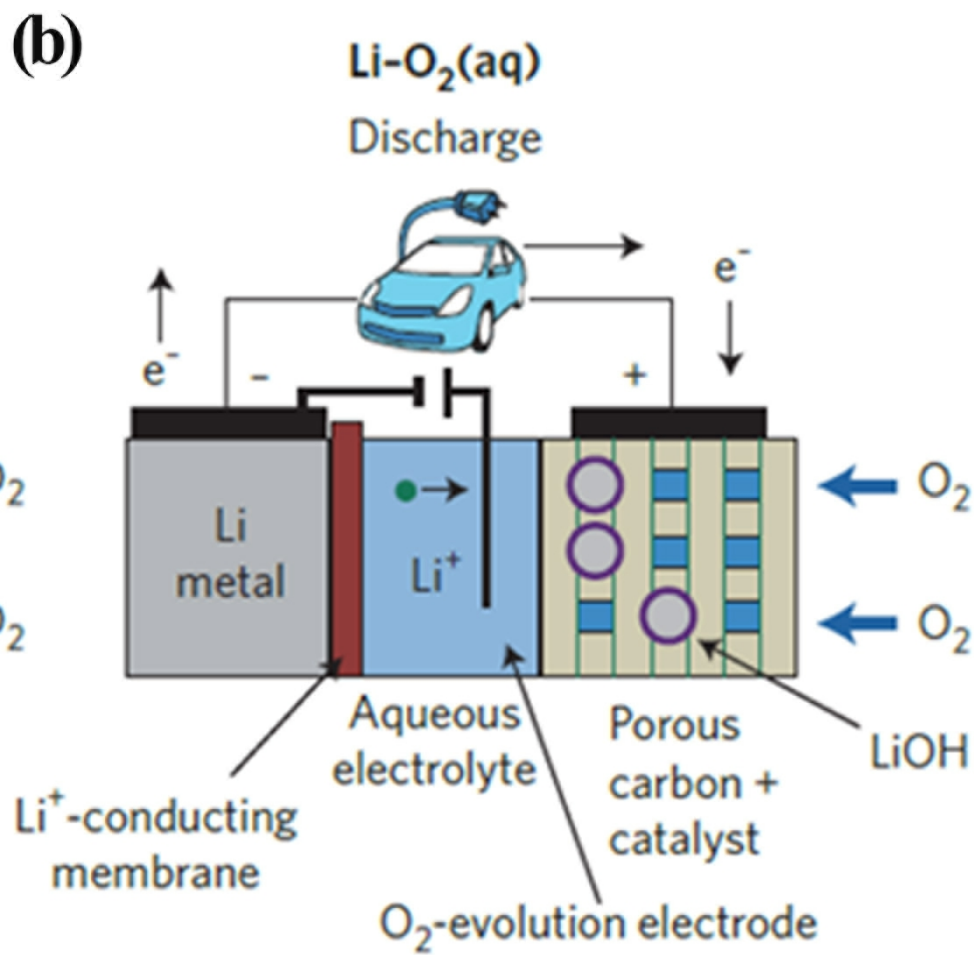

$$2Li^+ + 2e^- + O_2 \leftrightarrow Li_2O_2$$

$$2Li^+ + 2e^- + \tfrac{1}{2} O_2 + H_2O \leftrightarrow 2LiOH$$

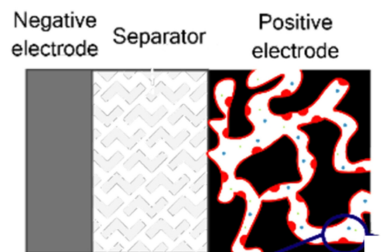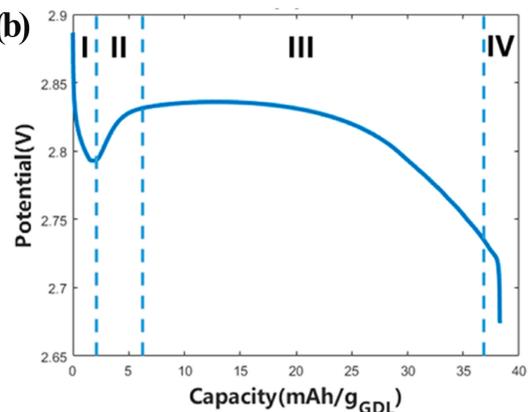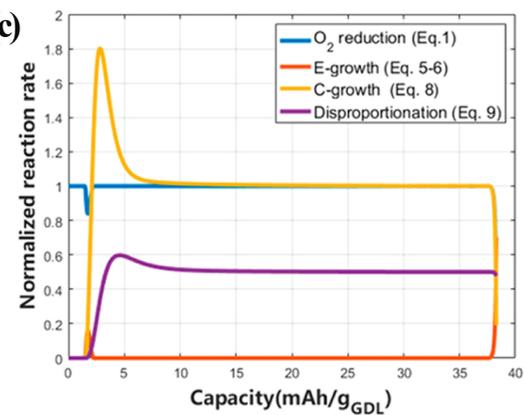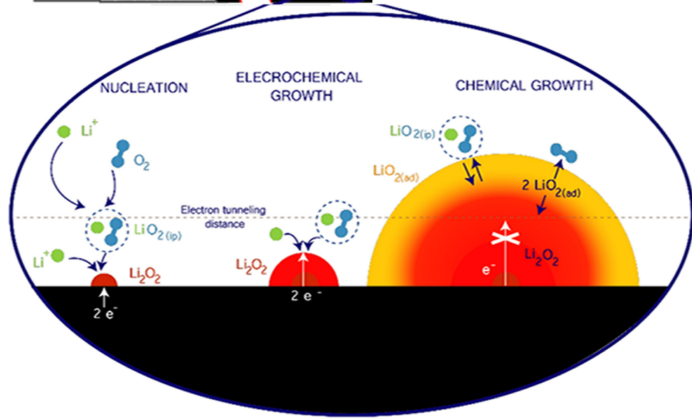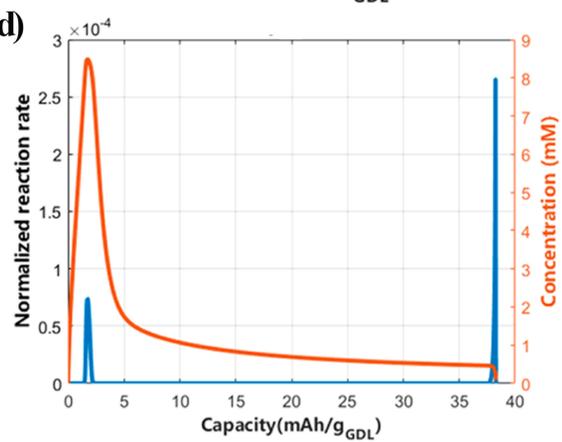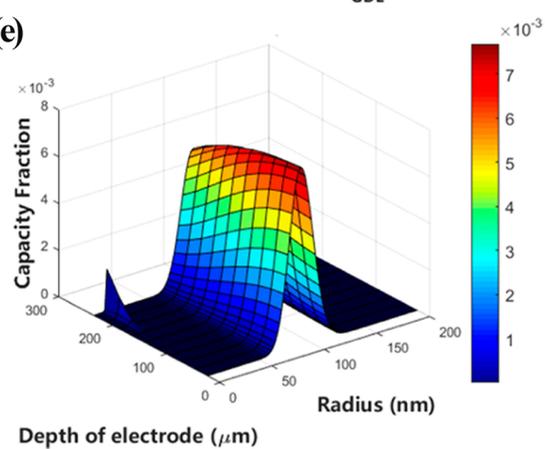

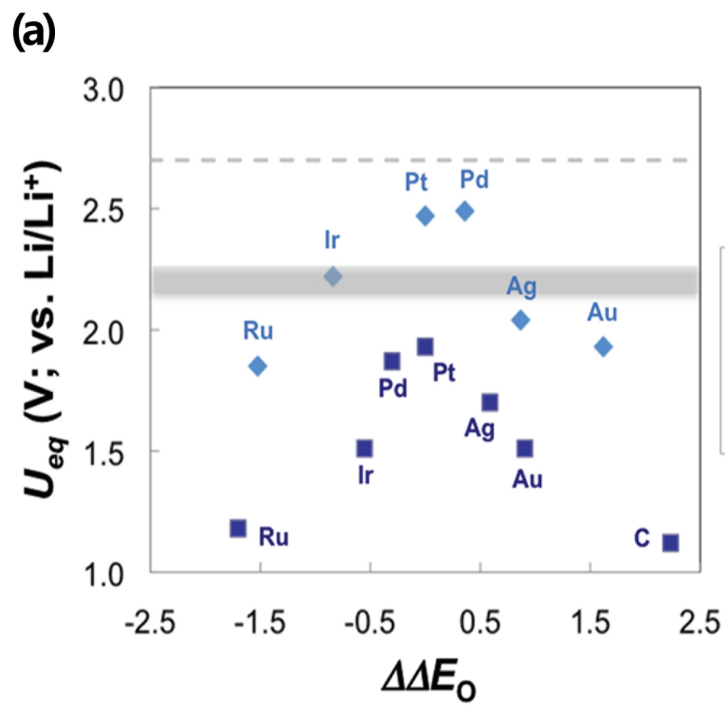 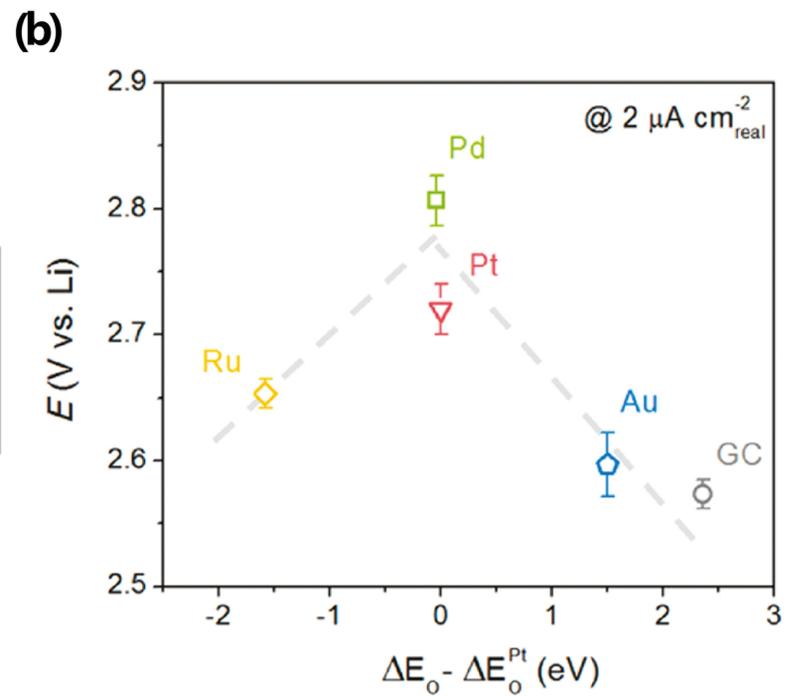

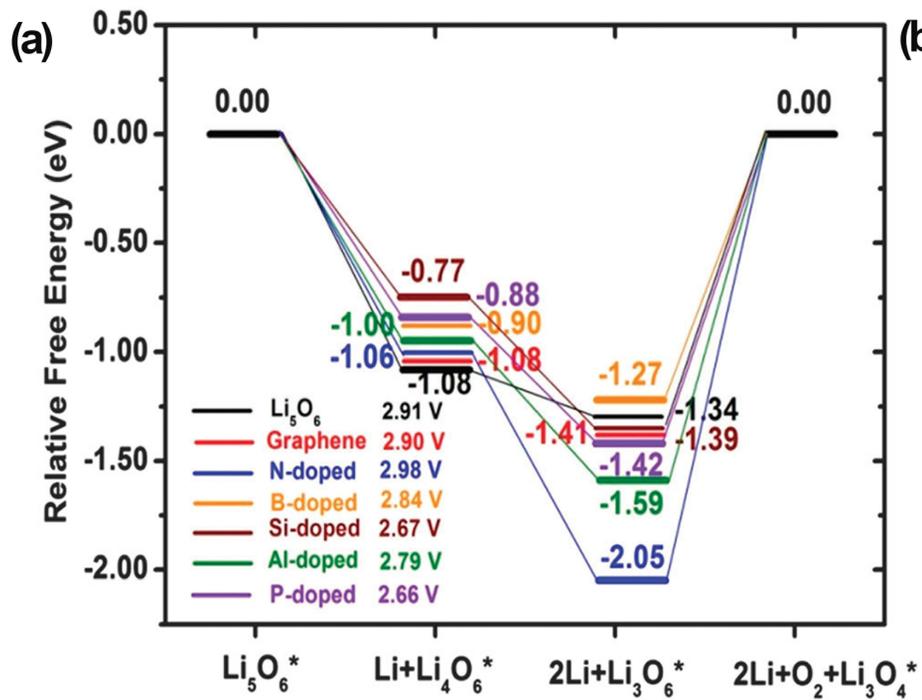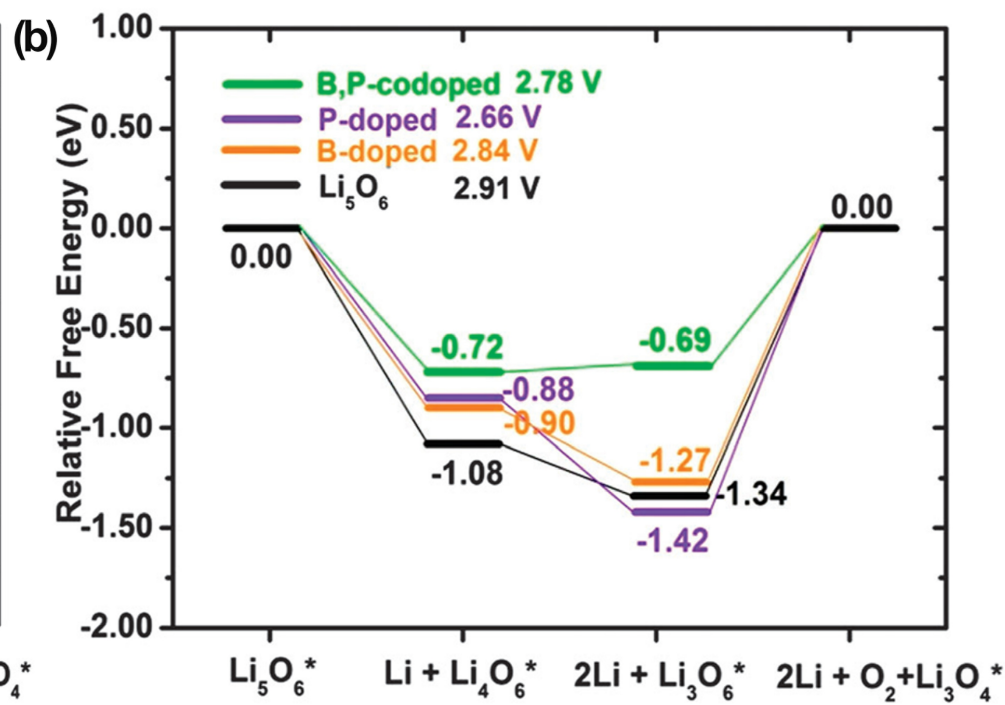

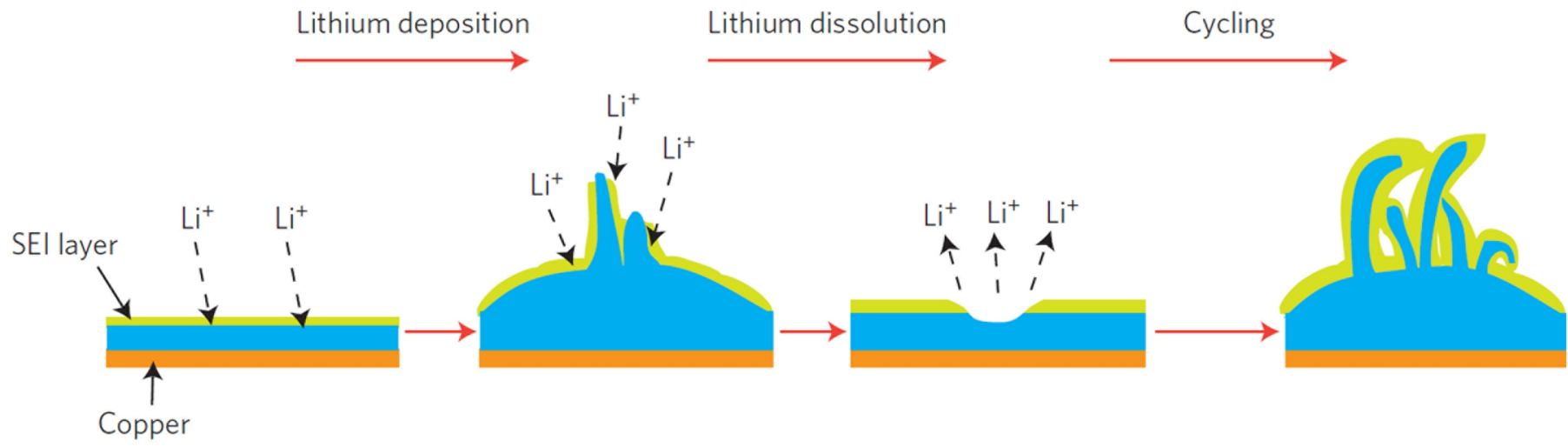

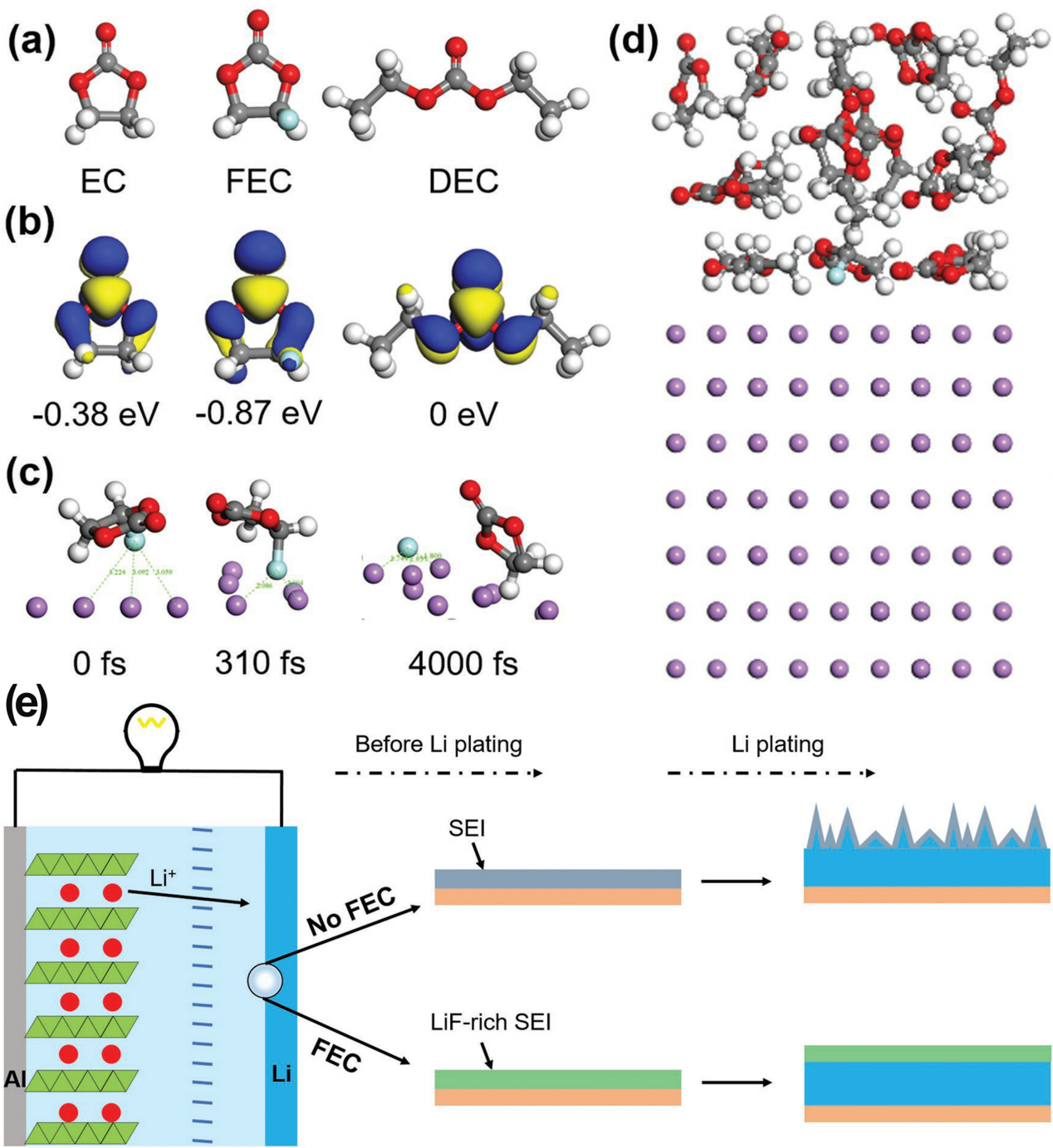

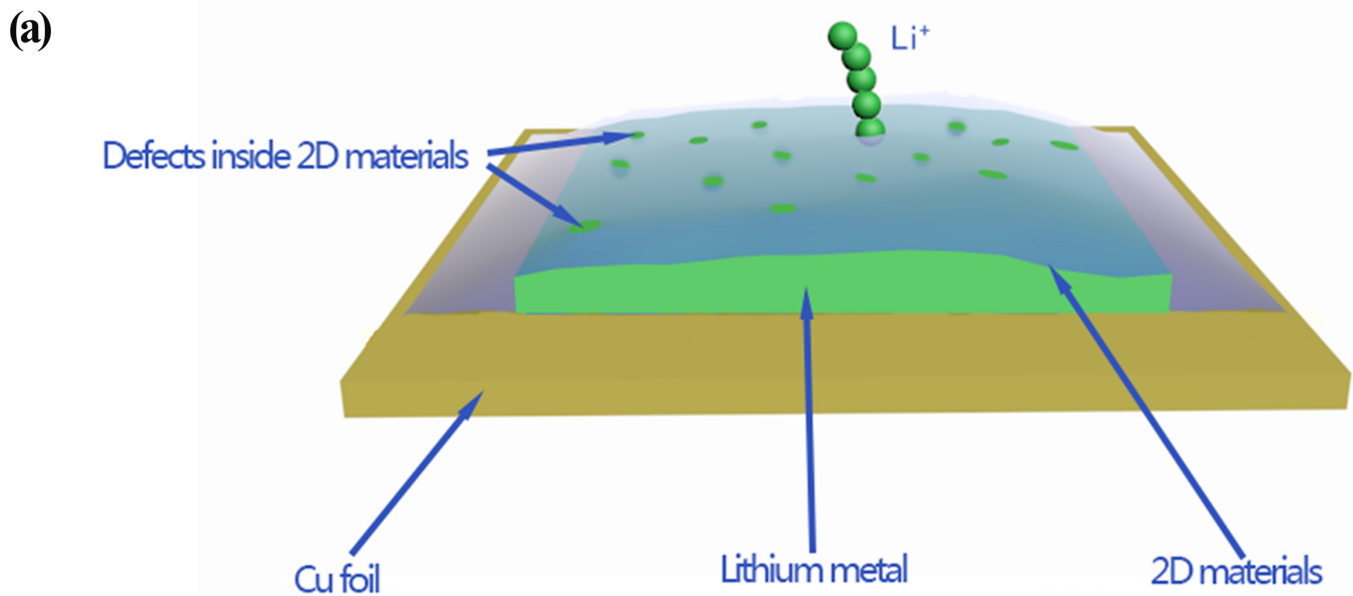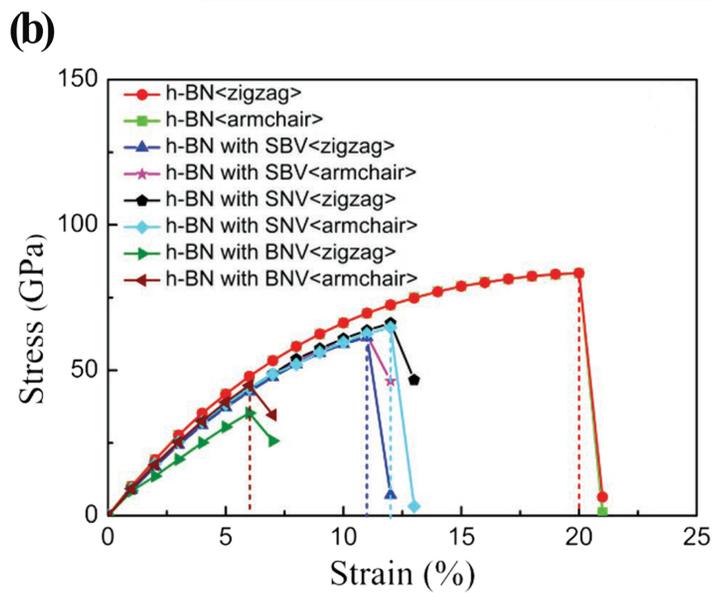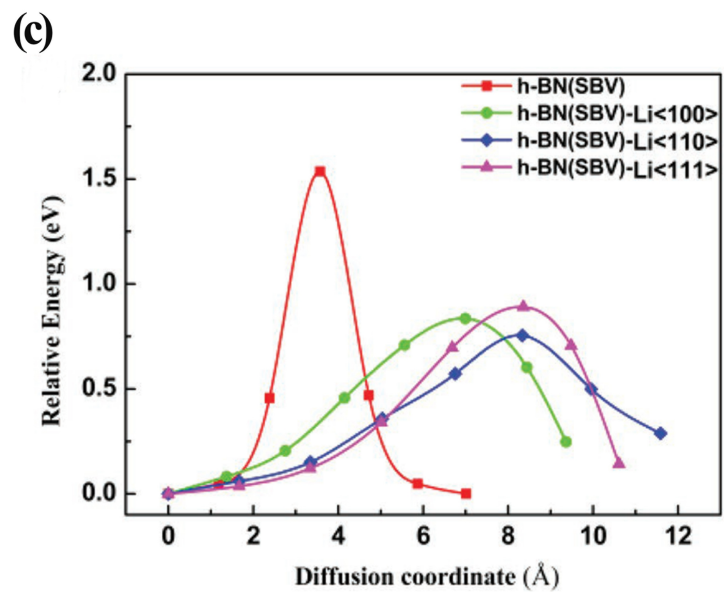

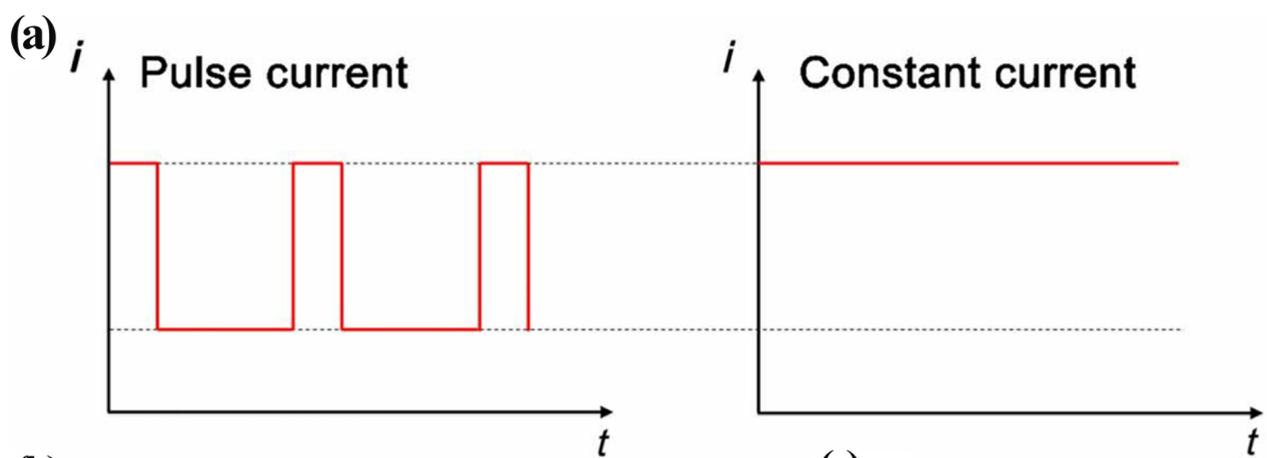
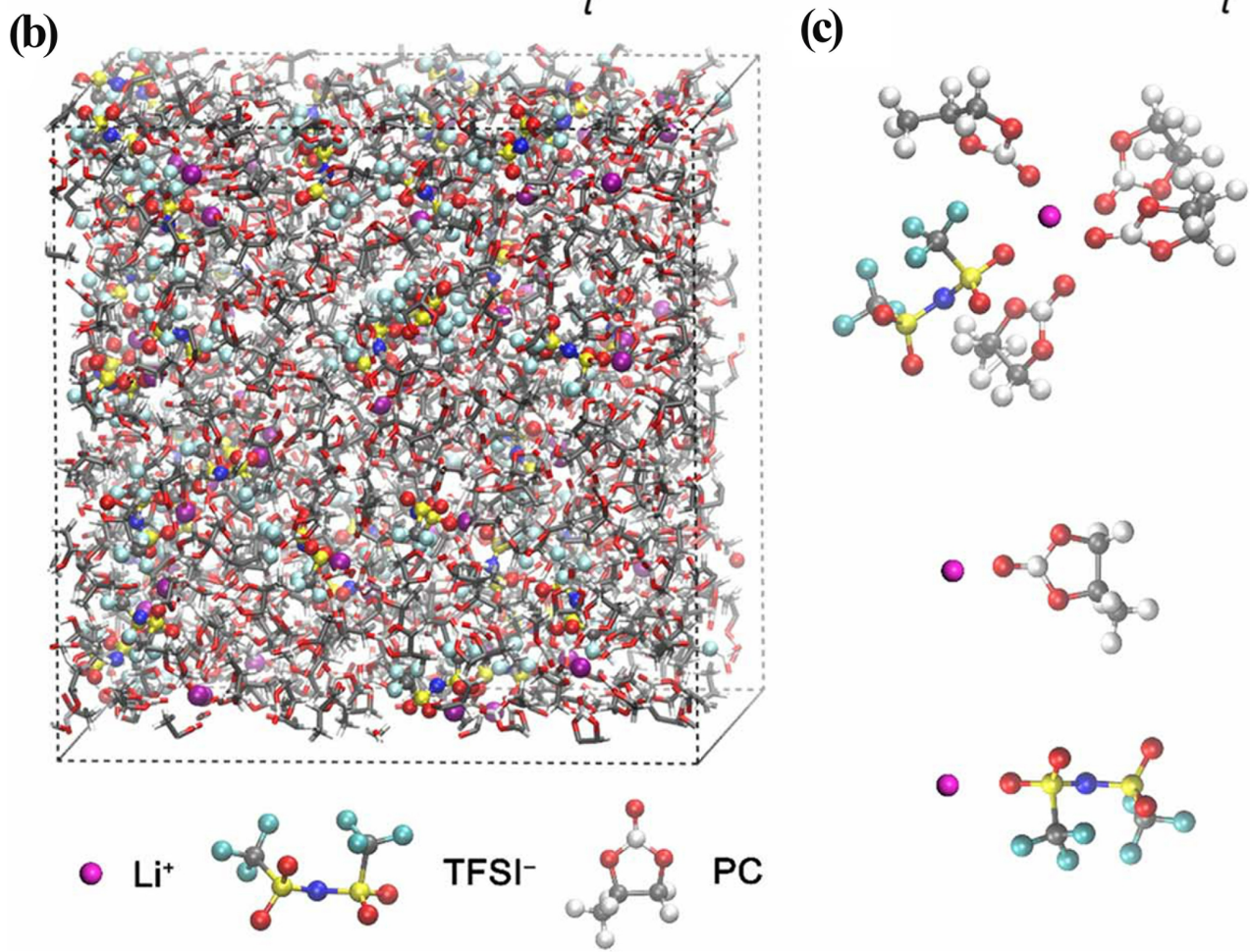

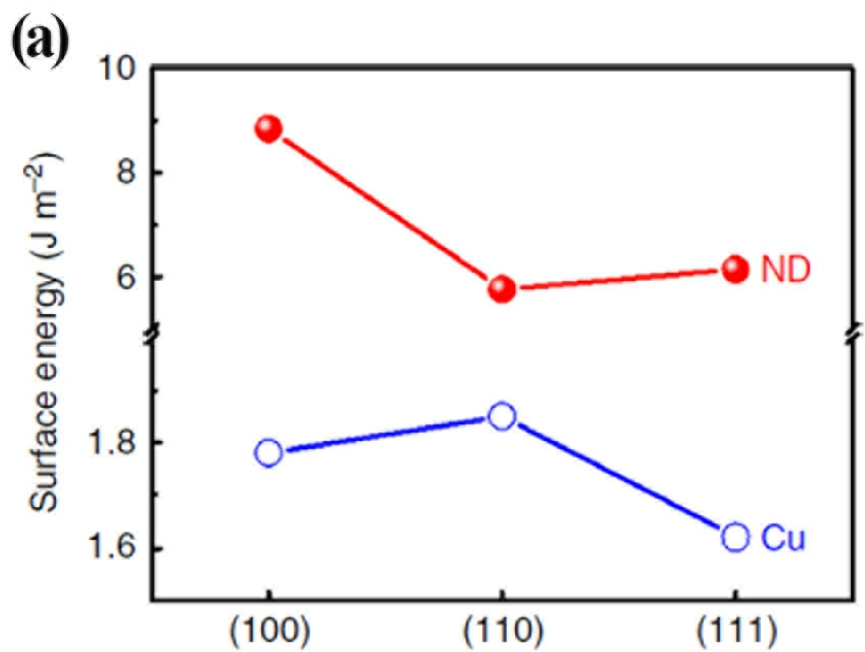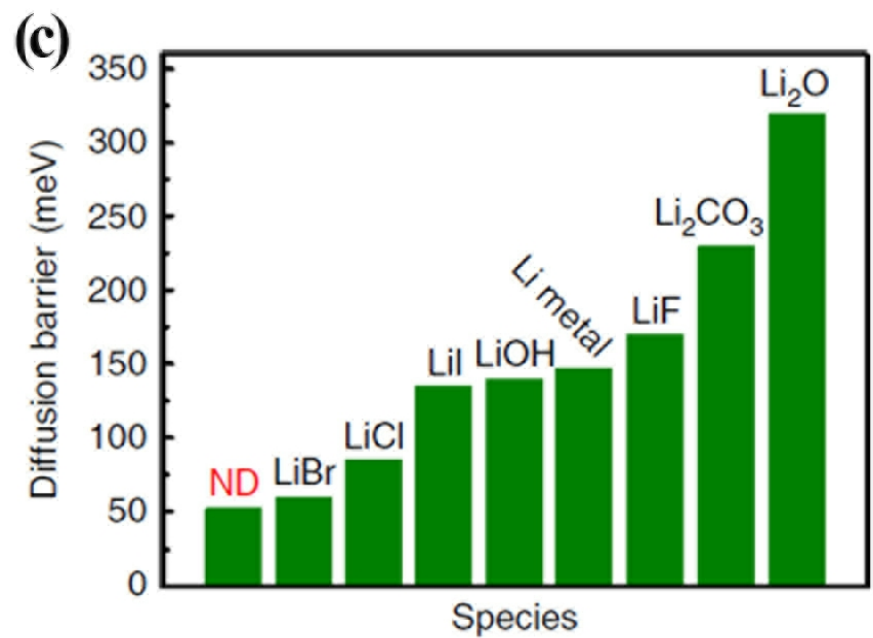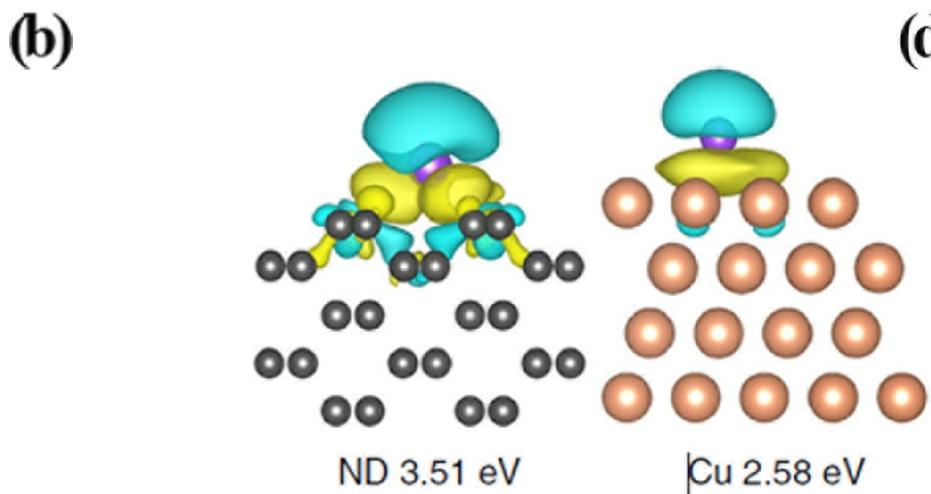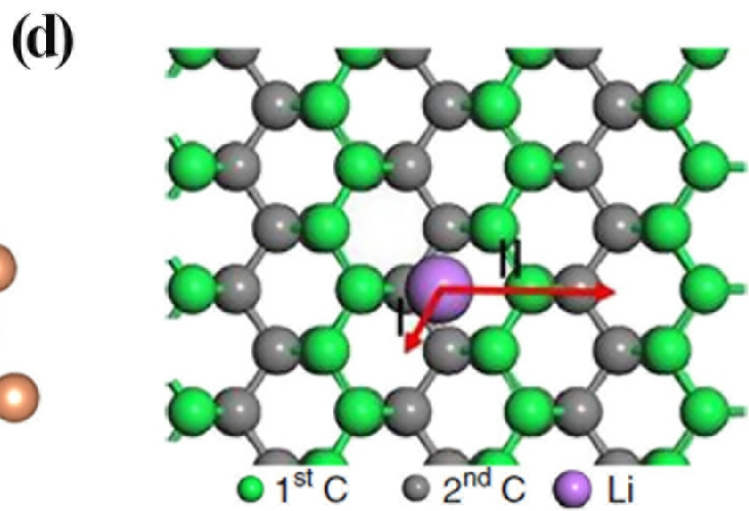

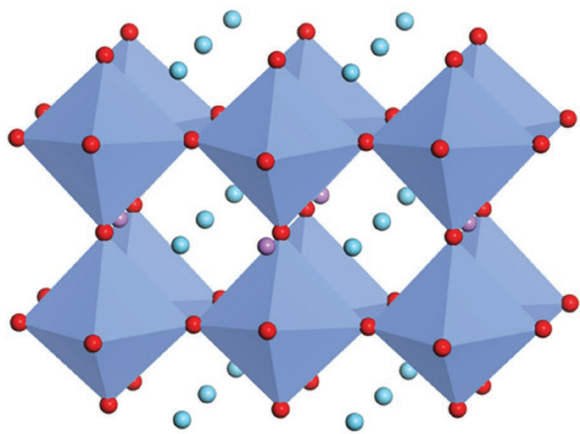
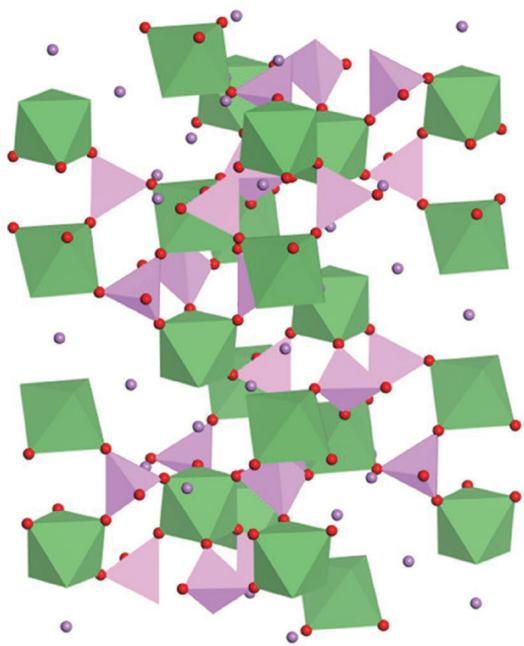
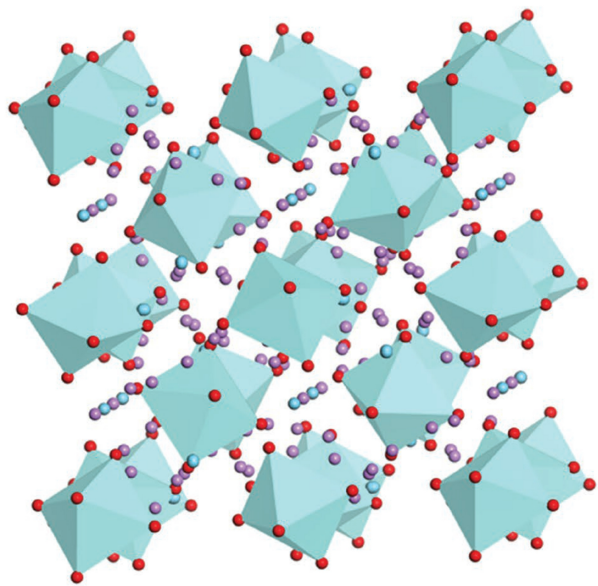
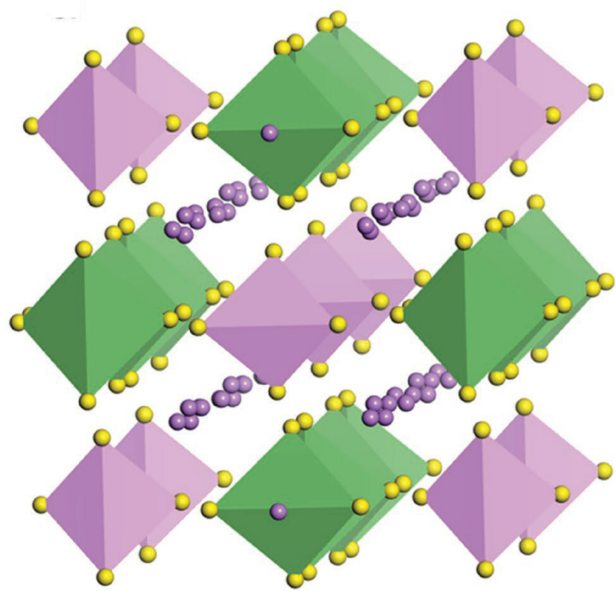

○ : O   ● : Li   ● : S   ● : La

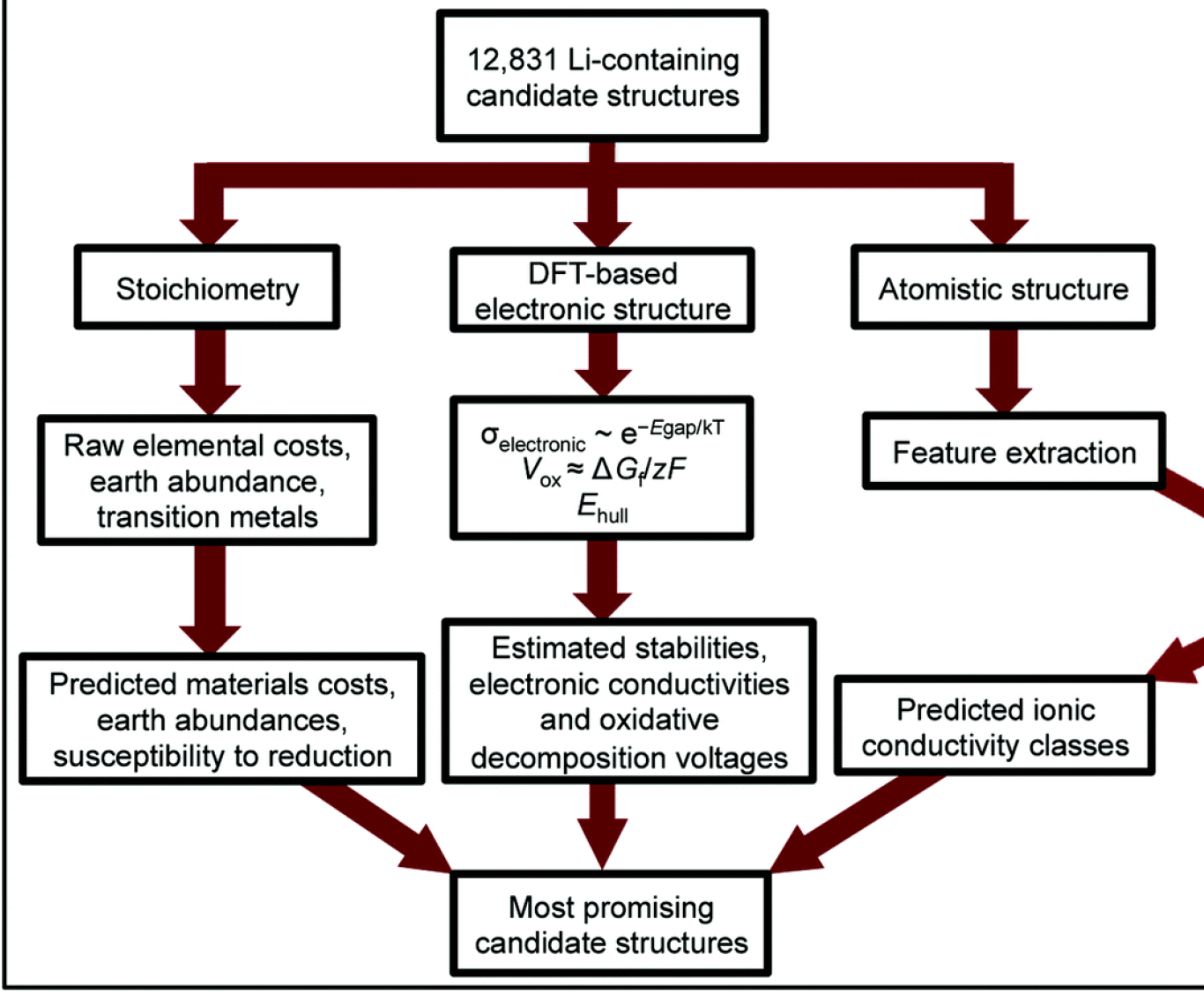

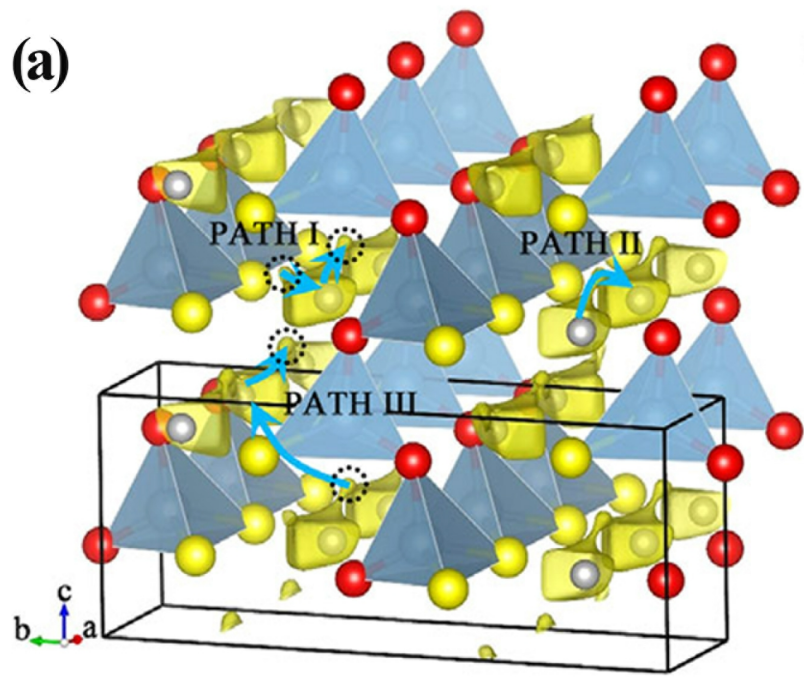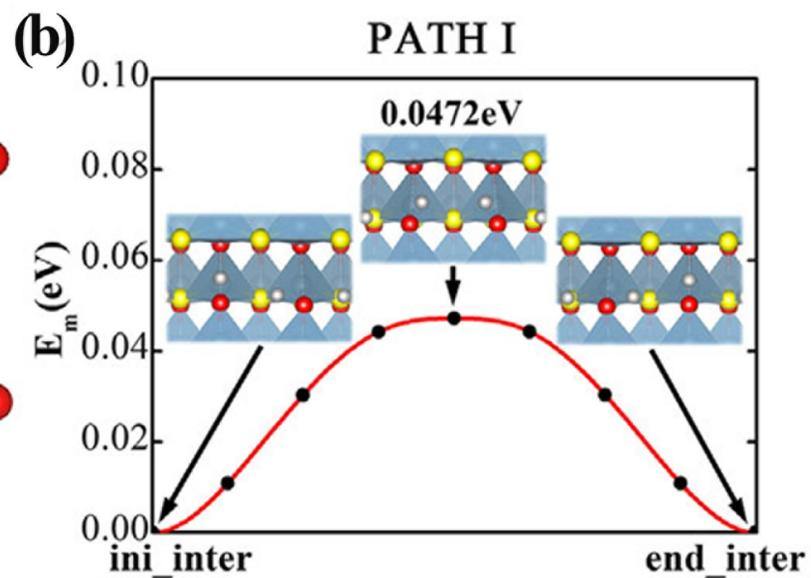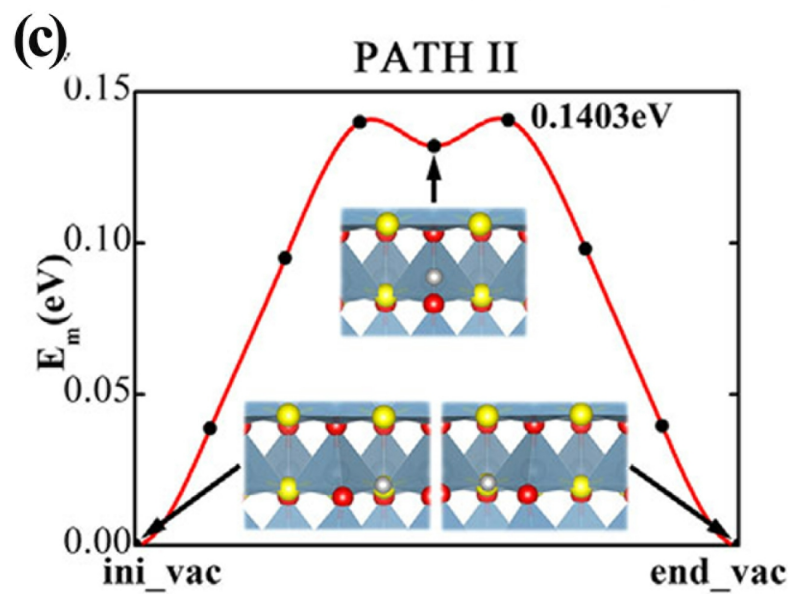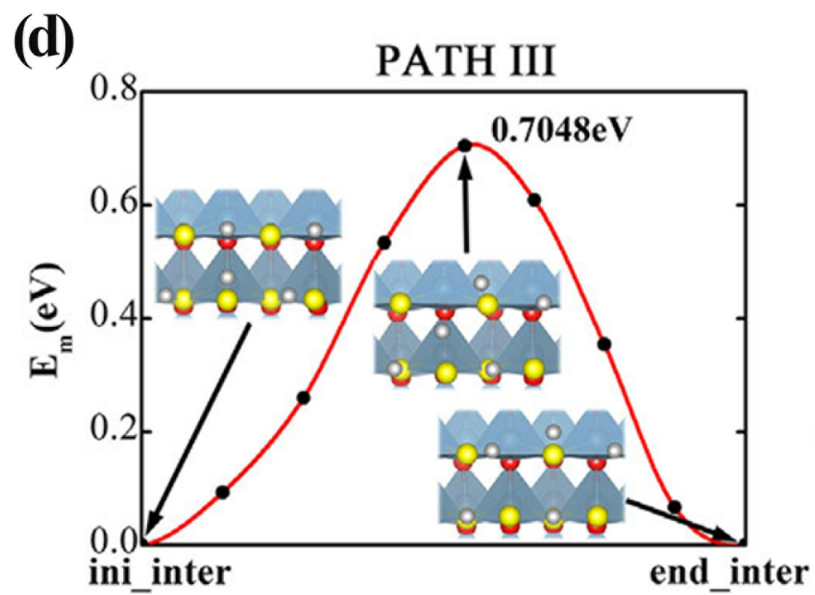

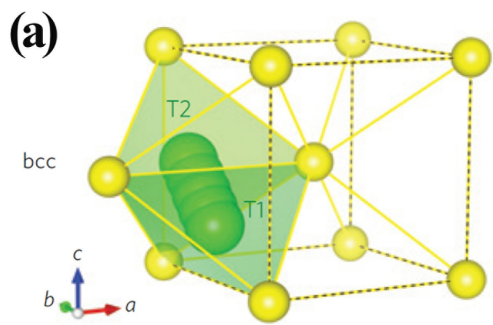
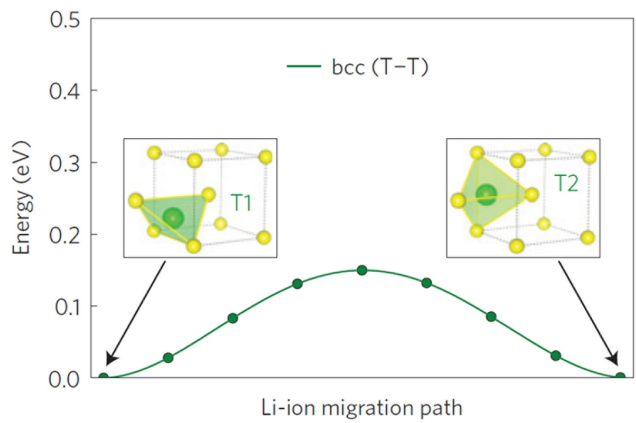
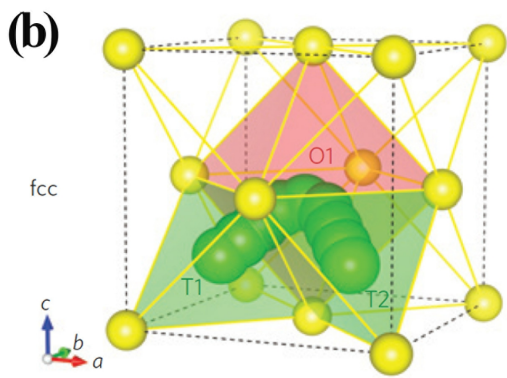
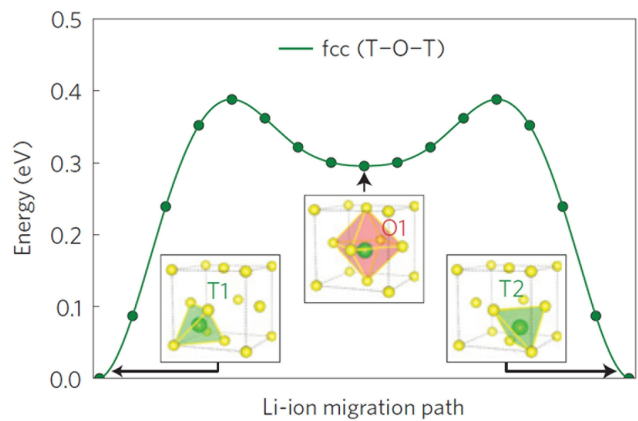
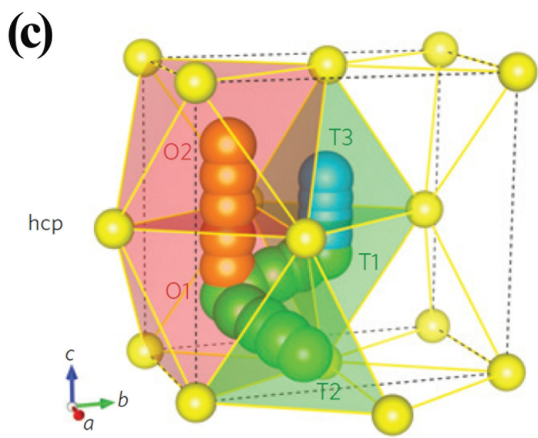
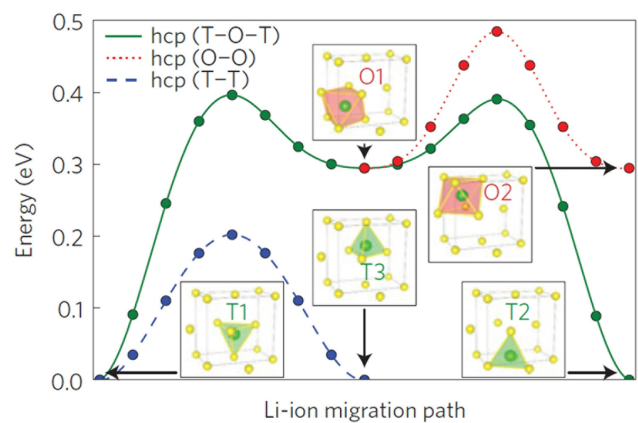

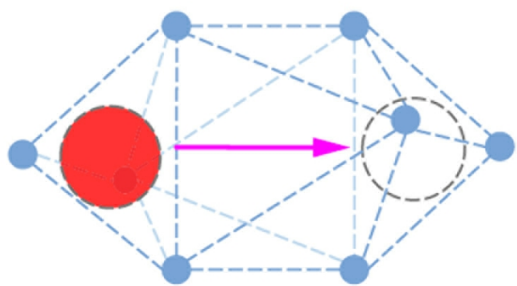
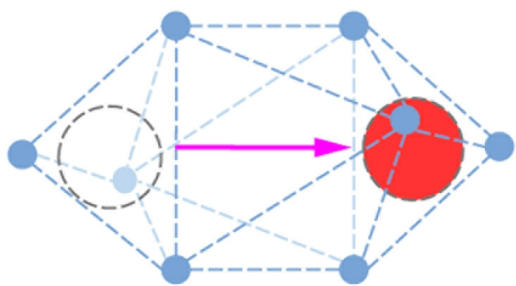

Energy landscape

Single-ion migration

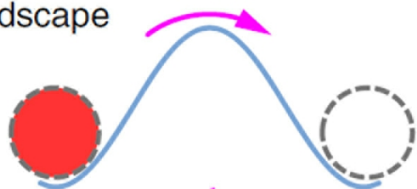
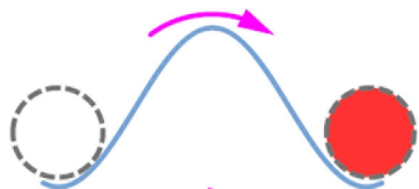

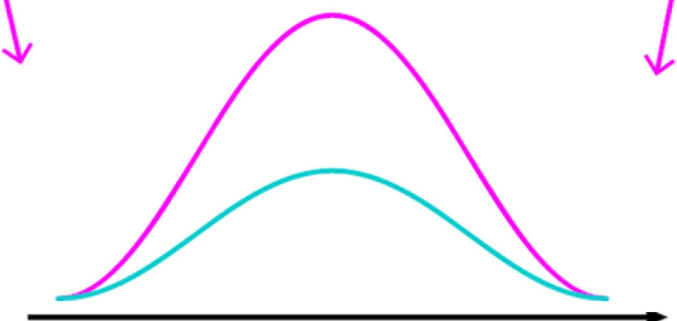

Energy barrier

Migration path

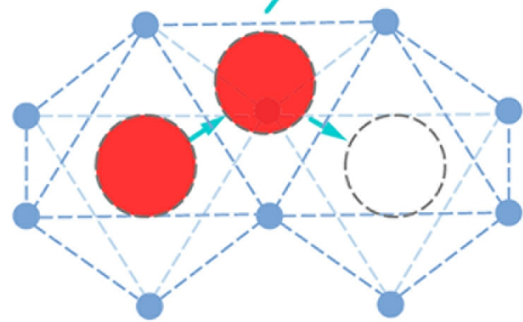
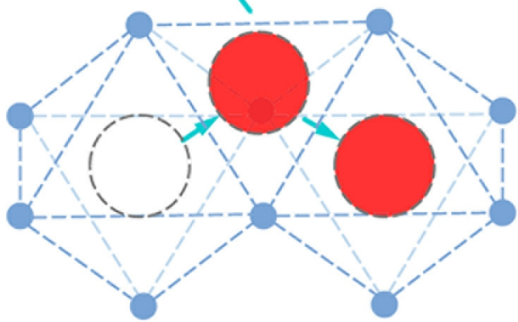

Energy landscape

Multi-ion concerted migration

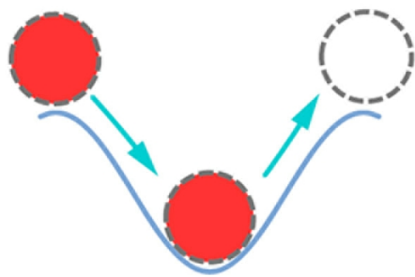
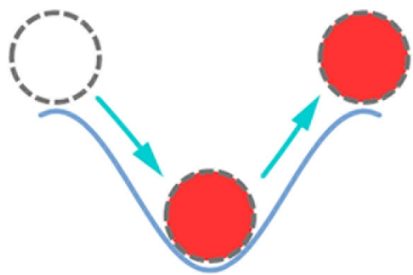

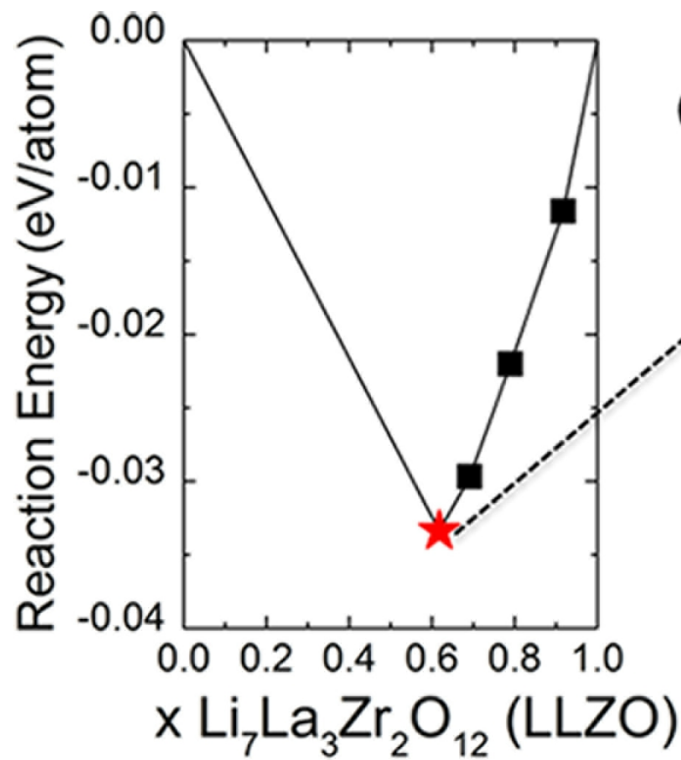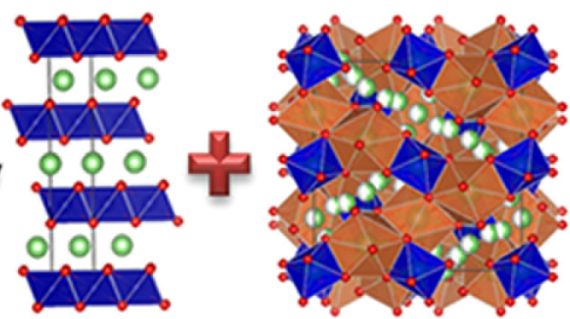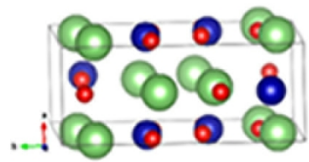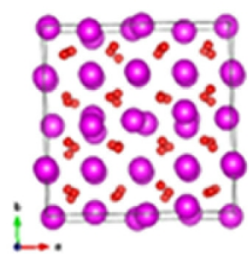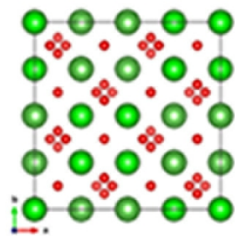